\documentclass[11pt,a4paper]{article}
\usepackage{jheppub}

\newcommand{\valpha}{v_\alpha}
\newcommand{\valphaeff}{v_\alpha^{\rm eff}}
\newcommand{\vmueff}{v_\mu^{\rm eff}}

\newcommand{\vmu}{v_\mu}

\newcommand{\vTbare}{v_{T,0}}

\def\msbar{$\overline{\hbox{MS}}$}

\newcommand{\seff}{\sin \theta_\text{eff}^\ell}

\usepackage{multirow}
\usepackage{caption}
\usepackage{subcaption}
\usepackage{placeins}

\title{Using the effective weak mixing angle as an input parameter in SMEFT}
\preprint{IPPP/23/77, MITP-23-074}

\author[a]{Anke Biek\"otter,}
\author[b]{Benjamin D.~Pecjak,}
\author[b]{Tommy Smith}
\affiliation[a]{PRISMA$^+$ Cluster of Excellence \& MITP, Johannes Gutenberg University, 55099 Mainz, Germany}
\affiliation[b]{Institute for Particle Physics Phenomenology, 
Durham University,  Durham DH1 3LE, UK}

\abstract{We implement electroweak renormalisation schemes involving the effective weak mixing angle
to NLO in Standard Model Effective Field Theory~(SMEFT).  After developing the necessary theoretical machinery, we analyse a select set
of electroweak precision observables in such input schemes. 
An attractive feature is that large corrections from top-quark loops appearing in other schemes
are absorbed into the definition of the effective weak mixing angle.  On the other hand, the renormalisation
condition which achieves this involves a large number of flavour-specific SMEFT couplings between
the $Z$ boson and charged leptons, motivating simple flavour assumptions such as minimal flavour
violation for practical applications.  The results of this paper provide a valuable new component for
estimating systematic uncertainties in SMEFT fits by performing analyses in multiple input schemes.
}

\begin{document}
\maketitle

\section{Introduction}
\label{sec:intro}

In the absence of direct new physics discoveries, Standard Model Effective Field Theory~(SMEFT)~\cite{Buchmuller:1985jz,Wilczek:1977pj,Grzadkowski:2010es} enables to  
describe in a model-independent fashion small deviations from Standard Model~(SM) predictions.  SMEFT calculations can be systematically 
improved by including quantum corrections as well as higher-order terms in the expansion in the new physics scale $\Lambda$.  The investigation of next-to-leading order (NLO) corrections in dimension-six SMEFT has been a focus of recent studies. While NLO QCD corrections have been fully automated~\cite{Degrande:2020evl}, NLO electroweak~(EW) corrections and, in a few instances next-to-next-to-leading order~(NNLO) QCD corrections, have been
calculated on a case-by-case basis for specific processes~\cite{Zhang:2013xya,Crivellin:2013hpa,Zhang:2014rja,Pruna:2014asa,Grober:2015cwa,Hartmann:2015oia,Ghezzi:2015vva,Hartmann:2015aia, Gauld:2015lmb, Aebischer:2015fzz,  Zhang:2016omx,BessidskaiaBylund:2016jvp, Maltoni:2016yxb, Gauld:2016kuu, Degrande:2016dqg,Hartmann:2016pil,Grazzini:2016paz,deFlorian:2017qfk,Deutschmann:2017qum,Baglio:2017bfe,Dawson:2018pyl,Degrande:2018fog,Vryonidou:2018eyv,Dedes:2018seb,Grazzini:2018eyk,Dawson:2018liq,Dawson:2018jlg,Dawson:2018dxp,Neumann:2019kvk,Dedes:2019bew, Cullen:2019nnr, Boughezal:2019xpp, Dawson:2019clf,Baglio:2019uty,Haisch:2020ahr, Cullen:2020zof, David:2020pzt, Dittmaier:2021fls,Dawson:2021ofa,Boughezal:2021tih,Battaglia:2021nys,Kley:2021yhn,Faham:2021zet,Haisch:2022nwz,Heinrich:2022idm,Bhardwaj:2022qtk,Asteriadis:2022ras,Bellafronte:2023amz,Kidonakis:2023htm,Gauld:2023gtb,Heinrich:2023rsd}. 
 
 An essential ingredient common to all NLO calculations is the choice of the EW input scheme.
In the recent work~\cite{Biekotter:2023xle},   we systematically examined at NLO in SMEFT the commonly employed $\alpha$, $\alpha_\mu$, and LEP schemes, which are defined in table~\ref{tab:schemeDef} and use combinations of the Fermi constant $G_F$, the masses of the $W$ and $Z$ bosons,  $M_W$ and $M_Z$, as well as the electromagnetic coupling $\alpha$ as the three independent EW input parameters.
This involved cataloguing the  set of Wilson coefficients entering finite parts of observables after the cancellation of UV divergences in different schemes, identifying  dominant sets of universal corrections associated with contributions from top-quark loops, 
and providing a methodology for including these scheme-dependent universal NLO corrections in the LO results, thus extending 
previous discussions of EW input schemes in SMEFT \cite{Brivio:2017bnu, Brivio:2021yjb}. 

In the SM, several studies have proposed EW input schemes which use the effective leptonic weak mixing angle $\seff$ as an input parameter~\cite{Kennedy:1988sn,Renard:1994ay,Ferroglia:2001cr,Ferroglia:2002rg,Chiesa:2019nqb,Amoroso:2023uux}. 
The effective leptonic weak mixing angle has been measured with per-mille level precision at LEP~\cite{ALEPH:2005ab}, the Tevatron~\cite{CDF:2018cnj} and the LHC~\cite{ATLAS:2015ihy,ATLAS:2018gqq,CMS:2018ktx,LHCb:2015jyu}. 
Its numerical precision is (more than an order of magnitude) below that of other commonly used input values such as the mass of the $W$ boson. 
Future experiments, such as the P2 experiment at MESA~\cite{Berger:2015aaa}, as well as the M{\o}ller~\cite{MOLLER:2014iki} and SoLID~\cite{Chen:2014psa,JeffersonLabSoLID:2022iod} experiments at Jefferson Laboratory,  will test this quantity with similar precision at lower energy scales.

In spite of the past and recent interest in EW input schemes involving $\seff$, a discussion in the context of SMEFT has not yet appeared in the literature. The aim of this paper is to fill this gap by incorporating  the $\{G_F,\seff,M_Z\}$ and $\{\alpha, \seff, M_Z\}$  input schemes 
into the methodology  of \cite{Biekotter:2023xle}.  We find that an attractive feature of these schemes 
is that large corrections from top-quark loops appearing in other schemes
are absorbed into the definition of the effective weak mixing angle.  On the other hand, the renormalisation
condition which achieves this involves a large number of flavour-specific SMEFT couplings between
the $Z$ boson and charged leptons, motivating simple flavour assumptions such as minimal flavour
violation for practical applications.  

We structure the discussion as follows.  In section~\ref{sec:seff} we cover 
the construction of UV counterterms in these schemes,  assembling and calculating the ingredients needed to implement them into 
NLO calculations in dimension-6 SMEFT.   Then, in section~\ref{sec:Numerics} we present a short 
study of numerical results for a select set of electroweak precision observables, including comparisons between all five EW input schemes listed in table~\ref{tab:schemeDef} at NLO in the SM and SMEFT, before concluding in section~\ref{sec:conclusions}.  In addition, some explicit results
for SMEFT expansion coefficients for derived quantities such as the $W$-boson mass are given in appendix~\ref{sec:Derived_SMEFT},
and a short description of minimal flavour violation is provided in appendix~\ref{app:flav_assumptions}.

The discussion in this paper is focused on providing the building blocks needed for NLO SMEFT calculations in  the input schemes involving $\seff$. 
Of course, an essential part of the actual implementation is calculating the loop diagrams (and real emission corrections) needed in 
the renormalisation process and its application to specific observables, which even for the modest set of processes considered in this
work involves a large number of Feynman diagrams and dozens of dimension-6 operators.
We have carried out these calculations using an in-house \texttt{FeynRules}~\cite{Alloul:2013bka} implementation of the SMEFT
Lagrangian and cross-checked our results with \texttt{SMEFTsim}~\cite{Brivio:2017btx,Brivio:2020onw}. 
For the calculation of matrix elements we employed \texttt{FeynArts} and~\texttt{FormCalc}~\cite{Hahn:2016ebn,Hahn:1998yk,Hahn:2000kx} and we extracted analytic results for Feynman integrals with \texttt{PackageX}~\cite{Patel:2015tea}.  
Numerical results were obtained  with \texttt{LoopTools}~\cite{Hahn:1998yk}. 
Analytic results for the observables in the $v_\mu^{\rm eff}$ and $v_\alpha^{\rm eff}$ schemes studied in section~\ref{sec:Numerics} are 
provided as electronic files with the arXiv submission of this work.

\begin{table}
	\centering
	\begin{tabular}{c | l }
	 scheme & inputs \\ \hline
		 $v_\mu^{\rm eff}$ & $G_F$, $\seff$, $M_Z$  \\ 
		
		$v_\alpha^{\rm eff}$ &  $\alpha$, $\seff$, $M_Z$\\ 
		
		 $\alpha_\mu$ &  $G_F$, $M_W$, $M_Z$\\
		
		 $\alpha$ &  $\alpha$, $M_W$, $M_Z$\\
		
		 LEP &  $G_F$, $\alpha$, $M_Z$
		 
	\end{tabular}
	\caption{\label{tab:schemeDef} Nomenclature for the EW input schemes considered in this work.}
\end{table}

\section{Using $\seff$ as an input parameter}
\label{sec:seff}
In this section we discuss renormalisation in two EW input schemes involving $\seff$:
\begin{itemize}
\item[(1)] The ``$v_\mu^{\rm eff}$ scheme'', which uses as inputs $\{G_F,\seff,M_Z\}$, where $G_F$ is the 
Fermi constant as measured in muon decay and $M_Z$ is on the on-shell $Z$-boson mass.
\item[(2)]The ``$v_\alpha^{\rm eff}$ scheme'', which uses as inputs $\{\alpha,\seff,M_Z\}$, where $\alpha$ is
the fine-structure constant renormalised in a given scheme.
\end{itemize}
In what follows, we shall refer to these two schemes collectively as the ``$v_\sigma^{\rm eff}$ schemes'', 
where the choice $\sigma\in\{\alpha,\mu\}$ can be used to select between them.

Our treatment of the $v_\sigma^{\rm eff}$ schemes follows closely the notation and procedures introduced in \cite{Biekotter:2023xle}.
We write the SMEFT Lagrangian up to dimension six as 
\begin{align}
\mathcal{L} = \mathcal{L}^{(4)} + \mathcal{L}^{(6)} \, ; \quad \mathcal{L}^{(6)} = \sum_{i} C_i \,  Q_i \, ,
\end{align}
where $\mathcal{L}^{(4)}$ is the SM Lagrangian and $\mathcal{L}^{(6)}$ denotes the dimension-six Lagrangian consisting of the operators $Q_i$ in the Warsaw basis~\cite{Grzadkowski:2010es} and the corresponding Wilson coefficients $C_i$, 
which are inherently suppressed by two powers of the new physics scale $\Lambda$. 
We list the 59 independent operators, which generally carry flavour indices, in table~\ref{op59}. 
We expand all quantities up to linear order in the dimension-six Wilson coefficients throughout this work. 

In order to implement the $v_\sigma^{\rm eff}$ schemes in a unified notation, we first write the tree-level  
Lagrangian in terms of $\{v_{T}, M_W, M_{Z}\}$, where $v_T$ is the vacuum expectation value (vev) of the
 SU(2) doublet Higgs field $H$
 \begin{align}
 \langle H^\dagger H \rangle = \frac{v_T^2}{2} \, .
 \end{align}
 The renormalised Lagrangian is obtained by interpreting the tree-level quantities as bare ones which are replaced
by renormalised parameters plus counterterms in a particular scheme.
For the inputs common to the two schemes, we relate the bare and renormalised parameters according to
\begin{align}
\label{eq:CTS}
M_{Z,0} & = M_Z\left(1+\Delta M_Z \right) \, ,\nonumber \\
s_{w,0} & = \sqrt{1-c_{w,0}^2} = \seff \left(1+ \Delta s_w\right) \equiv s_w\left(1+ \Delta s_w\right) \, , 
\end{align}
where here and throughout the paper we indicate bare parameters with a subscript $0$, and $c_{w,0}=M_{W,0}/M_{Z,0}$.  
The quantities $\Delta M_Z$ and $\Delta s_w$
appearing on the right-hand side of the above equations are counterterms, which are calculated in a SMEFT expansion
in loops and operator dimension, including tadpoles in the FJ tadpole scheme~\cite{Fleischer:1980ub}.

It will often be convenient to work with the quantity 
\begin{align}
M_W^{\rm eff} \equiv c_w M_Z \, ,  \quad c_w = \sqrt{1-s_w^2} \, .
\end{align}
The relation between $M_W^{\rm eff}$ and  the bare mass can be derived using eq.~(\ref{eq:CTS}).  Writing
\begin{align}
\label{eq:CTS_MW}
M_{W,0} &= M_W^{\rm eff}\left(1+ \Delta M_W^{\rm eff} \right)  \, ,
\end{align}
one finds
\begin{align}
\label{eq:MWeff_CT_derived}
\Delta M_W^{\rm eff} = \Delta M_Z - \frac{s_w^2}{c_w^2}\left[\Delta s_w+ \Delta M_Z \Delta s_w + \frac{1}{2c_w^2}\left(\Delta s_w\right)^2\right] + \dots \, ,
\end{align}
where the $\dots$ indicates terms not needed to NLO in the dimension-6 SMEFT expansion.

In addition to the counterterms for $s_w$ and $M_Z$, we also need those for $v_T$.  In the $v_\mu^{\rm eff}$ scheme, 
one uses 
\begin{align}
\label{eq:dV2_mu}
\frac{1}{v_{T,0}^2} & =\frac{1}{v_\mu^2}\left(1-  \Delta v_\mu^{\rm eff}\right) \equiv  \frac{1}{v_\mu^2}\left(1- \Delta v_\mu \right)\, , 
\end{align}
while in the $v_\alpha^{\rm eff}$ scheme one has instead
\begin{align}
\label{eq:dV2_alpha}
\frac{1}{v_{T,0}^2} & =\frac{1}{(v_\alpha^{\rm eff})^2}\left(1-   \Delta v_\alpha^{\rm eff}\right)  \equiv \frac{1}{v^2_\alpha}\left(1-  \Delta v_\alpha \right)  \, ,
\end{align}
where we have defined
\begin{align}
\label{eq:VEV_defs}
v_\mu \equiv \left(\sqrt{2} G_F\right)^{-\frac{1}{2}} \,  , \qquad v_\alpha^{\rm eff}\equiv v_\alpha \equiv  \frac{2 M_W^{\rm eff}s_w}{\sqrt{4\pi \alpha} }\,. 
\end{align}

To streamline the notation needed for discussing the $v_\sigma^{\rm eff}$ schemes, our definitions above
suppress the superscript ``eff'' on all quantities except for the scheme names and $M_W^{\rm eff}$, in order to distinguish it from the
on-shell $W$-boson mass $M_W$.  It is important to emphasize, however, that the SMEFT expansion coefficients of 
$\Delta v_\alpha$ and $\Delta v_\mu$ are {\it not} identical  to those in the $\alpha$ and $\alpha_\mu$ input schemes defined in
table~\ref{tab:schemeDef} and  discussed in  \cite{Biekotter:2023xle}, where $M_W$ instead  of $s_w$  is used as an input parameter.  
In the following two subsections we discuss renormalisation in the $v_\sigma^{\rm eff}$ schemes to NLO in SMEFT.

\subsection{The $v_\mu^{\rm eff}$ scheme}
In this section we give results for the SMEFT expansion of the counterterms needed for renormalisation in the
$v_\mu^{\rm eff}$ scheme, structuring the discussion in such a way that most results in the $v_\alpha^{\rm eff}$ scheme
can be obtained by a simple set of substitutions.

We begin with the determination of the counterterm $\Delta s_w$.  To this end, consider the amplitude for $Z\to \ell \ell$ 
decay, where $\ell \equiv \ell^i \in\{e,\mu,\tau\}$.  We can write the bare amplitude to NLO in SMEFT in the form 
 \begin{align}
 \label{eq:Zll_amp}
 {\cal A}_{0}(Z\to \ell \ell) = {\cal N}_0\left[{\cal A}^{\ell}_{L, 0} S_L + {\cal A}^{\ell}_{R, 0} S_R  \right]+ \dots \, ,
 \end{align}
where we have introduced the spinor structures
\begin{align}
\label{eq:Sdef}
S_L = \left[\bar{u}(p_{\ell^-})\gamma^\nu P_L v(p_{\ell^+}) \right] \epsilon_\nu^*(p_Z) \, , \quad
S_R = \left[\bar{u}(p_{\ell^-})\gamma^\nu P_R v(p_{\ell^+}) \right] \epsilon_\nu^*(p_Z) \, ,
\end{align}
with $P_L = (1-\gamma_5)/2$ and $P_R = (1+\gamma_5)/2$. The ellipsis in eq.~(\ref{eq:Zll_amp}) refers to spinor
structures appearing beyond LO in the SMEFT expansion, which do not  interfere with those above in the limit of vanishing lepton masses, and   the overall factor ${\cal N}_0$ is defined by
\begin{align}
\label{eq:N0}
{\cal N}_0 = \frac{M_{Z,0}}{v_{T,0}}\left(1 - \frac{v_{T,0}^2}{4} C_{HD,0}\right)\left(1+ \delta_{{\rm QED}}\right) \, , 
\end{align}
where $\delta_{\rm QED}$ refers to QED corrections.

We can write the SMEFT expansion of the bare amplitudes as
\begin{align}
\label{eq:A_LR_bare}
{\cal A}^\ell_{L/R, 0} &=  
 {\cal A}_{L/R,0}^{(4,0)} +\vTbare^2 {\cal A}_{L/R,0}^{\ell(6,0)}+ \frac{1}{\vTbare^2}{\cal A}_{L/R,0}^{(4,1)} + {\cal A}_{L/R,0}^{\ell(6,1)} \, , 
\end{align}
where the superscript $(i,j)$ labels the operator dimension $i$ contribution to the $j$-loop diagram, and we have pulled
out explicit factors of $v_{T,0}$ such that the coefficients ${\cal A}^{(i,j)}$ do not depend on $v_{T,0}$.\footnote{An implicit 
dependence on $v_T$ in the $(6,1)$ coefficients occurs through the Class-1 coefficient $C_W$. }
The notation makes clear that the dimension-6 amplitudes depend on the lepton species $\ell$ while those in the SM do not.

Suppressing the subscript 0, the tree-level SM amplitudes read
\begin{align} 
 {\cal A}_{L}^{(4,0)} = -1 + 2 s_{w}^2 \equiv  -c_{2w}, \quad  {\cal A}_{R}^{(4,0)} = 2 s_{w}^2 \, ,
\end{align}
while in SMEFT
\begin{align}
 {\cal A}_{L/R}^{\ell(6,0)} & = G^{(6,0)} + g^{\ell(6,0)}_{L/R} \ ,
\end{align}
where the explicit results for decay into lepton species $\ell^i$ are  
\begin{align}
	\label{eq:gdef}
G^{(6,0)} & = - c_w^2 C_{HD} - 2 c_w s_w C_{HWB}  \, ,\nonumber \\
g^{\ell(6,0)}_L & = -   C_{\substack{Hl \\ ii}}^{(1)}  -  C_{\substack{Hl\\ ii}}^{(3)}\, ,\nonumber \\
g^{\ell(6,0)}_R & = -   C_{\substack{He\\ ii}} \,.
\end{align} 
 
 Consider now the definition of the effective weak mixing angle 
 \begin{align}
 \label{eq:sw_eff_def}
 \sin^2 \theta_{\rm eff}^\ell = - \frac{1}{2} \, {\rm Re} \, \left(\frac{G_R^\ell(M_Z^2)}{G_L^\ell(M_Z^2)  - G_R^\ell(M_Z^2) }\right) \, , 
 \end{align}
where the $G^\ell_{L,R}$ are experimentally measured form factors at $q^2=M_Z^2$~\cite{ALEPH:2005ab,CDF:2018cnj,ATLAS:2015ihy,ATLAS:2018gqq,CMS:2018ktx,LHCb:2015jyu}. 
The counterterm $\Delta s_w$ in eq.~(\ref{eq:CTS}) is determined to all orders in the SMEFT expansion through the 
renormalisation condition
\begin{align}
\label{eq:Ren_Cond_SW}
 \sin^2 \theta_{\rm eff}^\ell  = s_w^2 \ .
 \end{align}
To implement this renormalisation condition order by order in SMEFT, we first substitute the $G^{\ell}_L (G^{\ell}_R)$ in eq.~(\ref{eq:sw_eff_def}) with the coefficients  of $S_L (S_R)$ in eq.~(\ref{eq:Zll_amp}), 
 and replace the bare quantities with renormalised ones plus associated
 counterterms. We write the SMEFT expansions of $\Delta s_w$ and $\Delta v_\mu$ in the $v_\mu^{\rm eff}$ scheme  as
 \begin{align}
 \label{eq:sw_vmu_exp}
 \Delta s_w &= v_\mu^2 \Delta s_w^{(6,0)} + \frac{1}{\vmu^2} \Delta s_w^{(4,1)} + \Delta s_w^{(6,1,\mu)} \,, \nonumber\\
 \Delta v_\mu &= v_\mu^2 \Delta v_\mu^{(6,0)} + \frac{1}{\vmu^2} \Delta v_\mu^{(4,1)} + \Delta v_\mu^{(6,1)}  .
 \end{align}
 The superscripts $(6,0)$ and $(4,1)$ have the same meaning as in eq.~(\ref{eq:A_LR_bare}), while the coefficient
$\Delta s_w^{(6,1,\mu)}$ contains an extra superscript $\mu$ to indicate that $v_T$ has been renormalised as in
 eq.~(\ref{eq:dV2_mu}). Isolating the dependence on $\Delta v_\mu$ allows us to write
\begin{align}
  \label{eq:DSW61mu}
  \Delta s_w^{(6,1,\mu)} & =  \Delta s_w^{(6,1)} -\Delta s_w^{(4,1)} \Delta v_\mu^{(6,0)} +  \Delta s_w^{(6,0)} \Delta v_\mu^{(4,1)} \, ,
 \end{align}
where the coefficient $\Delta s_w^{(6,1)}$ does not depend on the renormalisation scheme for $v_T$.
 
The construction of renormalised $Z\to \ell\ell$ decay amplitudes also requires the on-shell wavefunction renormalisation factors
of the external $Z$-boson and lepton fields.  For the lepton fields, we can write the SMEFT expansion of the wavefunction renormalisation
factors as 
 \begin{align}
 \ell_{L/R,0} & = 
 \left[1+   \frac{1}{2v_{T,0}^2}\Delta Z_{L/R,0}^{\ell(4,1)} +  \frac{1}{2} \Delta Z_{L/R,0}^{\ell(6,1)} \right]\ell_{L/R} \nonumber \\
 & = \left[1+   \frac{1}{2\vmu^2}\Delta Z_{L/R}^{\ell(4,1)} + 
  \frac{1}{2} \left(\Delta Z_{L/R}^{\ell(6,1)}  - \Delta v_\mu^{(6,0)}\Delta Z_{L/R}^{\ell(4,1)} \right) \right] \ell_{L/R} \, . 
 \end{align} 
In the first line all terms are expressed in terms of the bare parameters $v_{T,0},s_{w,0}$, while in the second renormalised 
parameters are used. The notation is such that 
\begin{align}
\label{eq:Z61_ren}
\Delta Z_{L/R}^{\ell(4,1)} & = \Delta Z_{L/R,0}^{\ell(4,1)}\bigg|_{s_{w,0}\to s_w} \nonumber , \\
\Delta Z_{L/R}^{\ell(6,1)}&  =\Delta s_w^{(6,0)} s_w\frac{\partial}{\partial s_w} \Delta Z_{L/R}^{\ell(4,1)} +
 \Delta Z_{L/R,0}^{\ell(6,1)}    \bigg|_{s_{w,0}\to s_w}      \, , 
\end{align}
where the derivative in the SMEFT counterterm arises from replacing the implicit 
dependence on $s_{w,0}$ in the SM counterterm $\Delta Z_{L/R,0}^{\ell(4,1)}$ with the right-hand side of eq.~(\ref{eq:sw_vmu_exp}) and 
performing a SMEFT expansion. 
We emphasise that the $\Delta Z^{\ell}_{L/R}$ do {\it not} include QED corrections, which are instead contained in the
 factor $\delta_{\rm QED}$ in eq.~(\ref{eq:N0}).   Wavefunction renormalisation graphs related to the $Z$-boson two-point function can be absorbed into the factor ${\cal N}_0$ in eq.~(\ref{eq:Zll_amp}). Since ${\cal N}_0$ drops out of 
 the ratio in eq.~(\ref{eq:sw_eff_def}) we do not discuss these two terms further.  On the other hand, contributions from the $Z\gamma$ 
 two-point function, where $\gamma$ denotes the photon, are included in the definition of ${\cal A}_{L/R,0}^{\ell}$. 
 
 Performing a tree-level SMEFT expansion on eq.~(\ref{eq:sw_eff_def}) using the above equations yields 
 \begin{align}
 \label{eq:DSW60}
\Delta s_w^{(6,0)} & = 
       -\frac{1}{4 s_w^2}\left[G^{(6,0)} + 2 s_w^2 g_L^{\ell(6,0)} +  c_{2w}g_R^{\ell(6,0)}  \right] \nonumber\\
& = \frac{1}{4s_w^2}\left[c_w^2 C_{HD} + 2 c_w s_w C_{HWB}+ 2 s_w^2\left(C_{\substack{H l\\ ii}}^{(3)}  +  C_{\substack{H l\\ ii}}^{(1)}\right)  + c_{2w} C_{\substack{He\\ ii}}  \right] \,, 
 \end{align}
 while the one-loop result in the SM is 
  \begin{align}
 \label{eq:DSW41}
\Delta s_w^{(4,1)} & = 
       -\frac{1}{2}\, {\rm Re}\, \left[ {\cal A}_{L,0}^{(4,1)} + \frac{c_{2w}}{2s_w^2} {\cal A}_{R,0}^{(4,1)} 
       + c_{2w}\left(\Delta Z_R^{\ell(4,1)}- \Delta Z_L^{\ell(4,1)}\right) \right] \,. 
 \end{align}
The part of the one-loop SMEFT result which is independent of the renormalisation scheme for $v_T$ is
 \begin{align}
 \label{eq:DSW61}
 \Delta s_w^{(6,1)}  = & -\frac{1}{2}\,{\rm  Re} \, 
 \bigg\{{\cal A}_{L,0}^{(6,1)} + {\cal A}_{L}^{(6,0)} \Delta Z_L^{\ell(4,1)} +
 \frac{c_{2w}}{2s_w^2}\left({\cal A}_{R,0}^{(6,1)} + {\cal A}_{R}^{(6,0)} \Delta Z_R^{\ell(4,1)}\right)  
 \nonumber \\
 & +c_{2w}\left( \Delta Z_{R,0}^{\ell(6,1)}- \Delta Z_{L,0}^{\ell(6,1)}  \right) \bigg\}  + \frac{1}{2\epsilon} v_\mu^2 \Delta \dot{s}_w^{(6,0)}  
 \nonumber \\ &
 - \Delta s_w^{(4,1)}\left(\Delta s_w^{(6,0)} +\frac{1}{2} C_{HD} - \frac{c_{2w}}{2c_ws_w}C_{HWB}\right)
 \nonumber \\ &
 + \Delta s_w^{(6,0)}\left[ s_w \frac{\partial}{\partial s_w} \Delta s_w^{(4,1)} 
 - \frac{1}{2s_w^2}{\cal A}_{R,0}^{(4,1)}-  \Delta Z_R^{\ell(4,1)}  \right] \,.
 \,
 \end{align}
A couple of comments are in order concerning the form of this counterterm. First, the quantity $\Delta \dot{s}_w^{(6,0)}$ is obtained from $\Delta s_w^{(6,0)}$ through the replacement $C_i\to \dot{C}_i$,
where $\dot{C}_i\equiv d C_i/d\ln\mu$; the term involving this quantity arises from \msbar~renormalisation of the Wilson
coefficients in $d=4-2\epsilon$ dimensions, and the $\dot{C}_i$ were calculated at one loop in~\cite{Jenkins:2013zja,Jenkins:2013wua,Alonso:2013hga}.\footnote{The $\dot{C}_i$ typically depend on a large number of Wilson coefficients, so it is convenient to use the electronic implementation in \texttt{DsixTools}~\cite{Celis:2017hod, Fuentes-Martin:2020zaz}. As they are one-loop corrections they thus scale as $1/\vmu^2$ and so the counterterm is independent of $\vmu$.}  
Second, the final two lines are related to using eq.~(\ref{eq:sw_vmu_exp}) in the lower-order
amplitudes and then performing the SMEFT expansion.

To evaluate the full NLO SMEFT result $\Delta s_w^{(6,1,\mu)}$ in  eq.~(\ref{eq:DSW61mu}) requires also the 
counterterm $\Delta v_\mu$.  This counterterm, including tadpoles and without flavour assumptions,  
was determined at NLO in SMEFT in the $\alpha_\mu$ scheme in \cite{Biekotter:2023xle}, thus generalising the earlier result from~\cite{Dawson:2018pyl}.
Calling the expansion coefficients eq.~(\ref{eq:sw_vmu_exp}) in that scheme $\Delta v_\mu^{\alpha_\mu(i,j)}$, the relation with the 
present work is 
\begin{align}
\Delta v_\mu^{(6,0)} & =  \Delta v_\mu^{\alpha_\mu(6,0)} = C_{\substack{Hl \\ 11}}^{(3)} + C_{\substack{Hl \\ 22}}^{(3)}
  - C_{\substack{ll \\ 1221}} \, ,\nonumber \\
\Delta v_\mu^{(4,1)} & = \Delta v_\mu^{\alpha_\mu(4,1)}\bigg|_{M_W\to M^{\rm eff}_{W}} \, ,\nonumber \\
\Delta v_\mu^{(6,1)} & =  \Delta s_w^{(6,0)}  s_w\frac{\partial}{\partial s_w} \Delta v_\mu^{(4,1)} + \Delta v_\mu^{\alpha_\mu(6,1)} \bigg|_{M_W\to M^{\rm eff}_{W}}  \,.
\end{align}

It will be useful later on to have an expression for the large-$m_t$ limit of the loop corrections to the
counterterm $\Delta s_w$.  Here and below, the large-$m_t$ limit of a given quantity means the approximation 
where only terms proportional to positive powers of the top-quark mass $m_t$ in the limit $m_t\to \infty$ are kept.  
In the SM,  top-quark loops contribute to $\Delta s_w^{(4,1)}$ in eq.~(\ref{eq:DSW41}) only through the $Z\gamma$-mixing contribution to the bare
amplitudes ${\cal A}^{(4,1)}_{L/R,0}$.  It is easy to show, however, that this two-point function vanishes in the 
large-$m_t$ limit, so
\begin{align}
 \label{eq:DS41mu}
\Delta s^{(4,1)}_{w,t}  = 0 \,,
\end{align}
where here and below the subscript $t$ refers to the large-$m_t$ limit of a given quantity. 
In SMEFT, there are three contributions in the large-$m_t$ limit, which arise from $Z\gamma$ mixing, top-loop corrections from four-fermion operators, and a scheme-dependent correction proportional to $\Delta v_{\mu,t}^{(4,1)}$.  
The result can be written in the form
 \begin{align}
 \label{eq:DS61mu}
 \Delta s_{w,t}^{(6,1,\mu)} & = \Delta s_{w,t}^{Z\gamma {(6,1)}} +  \Delta s_{w,t}^{4f{(6,1)}} + \Delta s_w^{(6,0)}\Delta v_{\mu,t}^{(4,1,\mu)} + \dots \, , 
 \end{align}
 where the $\dots$ refers to divergent and tadpole contributions, which drop out of physical observables.  An explicit 
 calculation yields
 \begin{align}
  \Delta s_{w,t}^{Z\gamma{(6,1)}}& = \frac{\sqrt{2}c_w}{3 s_w}\frac{M_Z}{m_t}\Delta\rho_t^{(4,1)}
  \left[ c_w C_{\substack{uB\\33}}\left(-3+16s_w^2\right) + s_w C_{\substack{uW\\33}}\left(-11+16s_w^2\right)\right] \ln \left(\frac{\mu^2}{m_t^2}\right)   \, , \nonumber \\
 \Delta s_{w,t}^{4f{(6,1)}} & =  \Delta \rho_t^{(4,1)} 
 \left[C^{(3)}_{\substack{lq\\ii33}} - C^{(1)}_{\substack{lq\\ii33}}  +C_{\substack{lu\\ii33}} 
 +\frac{c_{2w}}{2s_w^2}\left(C_{\substack{eu\\ii33}}- C_{\substack{qe\\33ii}}   \right)\right] \ln \left(\frac{\mu^2}{m_t^2}\right)  \, ,\nonumber \\
 \Delta v_{\mu,t}^{(4,1)} & = -\Delta \rho_t^{(4,1)}\left[1 + 2 \ln \left(\frac{\mu^2}{m_t^2}\right)   \right] \, ,
 \end{align}
 where as above we omitted divergent and tadpole contributions, and quoted the results 
 in units of 
 \begin{align}
 \Delta \rho_t^{(4,1)} & \equiv \frac{3}{16\pi^2} m_t^2 \, .
 \end{align}
 
 \subsection{The $v_\alpha^{\rm eff}$ scheme}

The  $v_\alpha^{\rm eff}$ scheme differs from the  $v_\mu^{\rm eff}$ scheme through the renormalisation of $v_T$, which is performed
as in eq.~(\ref{eq:dV2_alpha}).  The SMEFT expansion coefficients of that counterterm, as well as those of $\Delta s_w$ in this scheme, are 
defined as in eq.~(\ref{eq:sw_vmu_exp}) after the replacement $\mu \to \alpha$. 

In order to calculate the expansion coefficients of $\Delta v_\alpha$, we will need those for $M_Z$ and electric charge renormalisation.
We define these as 
\begin{align}
\label{eq:MZ_e_ren} 
M_{Z,0} & = M_Z\left(1 + \frac{1}{\valpha^2}\Delta M_Z^{(4,1)} + \Delta M_Z^{(6,1)}- \Delta v_\alpha^{(6,0)}\Delta M_Z^{(4,1)} \right) \, ,  \nonumber\\
e_0 & = e\left(1 + \frac{1}{\valpha^2}\Delta e^{(4,1)} + \Delta e^{(6,1)}- \Delta v_\alpha^{(6,0)}\Delta e^{(4,1)} \right) \, ,
\end{align}
where the coefficients with superscript $(6,1)$  are calculated as in eq.~(\ref{eq:Z61_ren}).  It will also be useful to work with expansion coefficients of the 
derived counterterm $\Delta M_W^{\rm eff}$.   We define these 
coefficients as 
\begin{align}
\label{eq:MWeff_bare}
M_{W,0}& =  M_W^{\rm eff}\left[1 + \valpha^2 \Delta M_W^{{\rm eff}(6,0)} +\frac{1}{\valpha^2}  \Delta M_W^{{\rm eff}(4,1)}  + \Delta M_W^{{\rm eff}(6,1,\alpha)} \right] \, .
\end{align}
Performing a SMEFT expansion on eq.~(\ref{eq:MWeff_CT_derived}) leads to 
\begin{align}
\label{eq:MWeff_coeffs}
\Delta M_W^{{\rm eff}(6,0)} & = -\frac{s_w^2}{c_w^2}\Delta s_w^{(6,0)} \, ,\nonumber \\
\Delta M_W^{{\rm eff}(4,1)} & = \Delta M_Z^{(4,1)}  - \frac{s_w^2}{c_w^2} \Delta s_w^{(4,1)} \, , \nonumber \\
\Delta M_W^{{\rm eff}(6,1,\alpha)}  &=\Delta M_W^{{\rm eff}(6,1)} -\Delta M_W^{{\rm eff}(4,1)}\Delta v_\alpha^{(6,0)}
+ \Delta M_W^{{\rm eff}(6,0)} \Delta v_\alpha^{(4,1)} \, ,
\end{align}
where 
\begin{align}
\Delta M_W^{{\rm eff}(6,1)} = \Delta M_Z^{(6,1)} -\frac{s_w^2}{c_w^2}\left[\Delta s_w^{(6,1)} +\Delta s_w^{(6,0)}  \Delta M_Z^{(4,1)}+ \frac{1}{c_w^2} \Delta s_w^{(6,0)}\Delta s_w^{(4,1)} \right]  \, .
\end{align}

With this notation at hand, we can present a compact result for  the expansion coefficients of $\Delta v_\alpha$.  They read
 \begin{align}
\Delta v_\alpha^{(6,0)} & =  \Delta v_\alpha^{\alpha(6,0)} -\frac{ 2c_{2w}}{s_w^2}\Delta M_W^{{\rm eff} (6,0)} \, ,
\nonumber
\\
\Delta v_\alpha^{(4,1)} & =\, 2\left(\Delta M_W^{{\rm eff}(4,1)} + \Delta s_w^{(4,1)} - \Delta e^{(4,1)}\right)\, ,
\nonumber \\
 \Delta v_\alpha^{(6,1)} & =\, 2\left(\Delta M_W^{{\rm eff}(6,1)} + \Delta s_w^{(6,1)} - \Delta e^{(6,1)}\right) 
-  \Delta v_\alpha^{(6,0)}  \Delta v_\alpha^{(4,1)}\nonumber  \\
&  + \frac{ 2}{c_w s_w} \left[C_{HWB}+\frac{c_w}{2s_w}C_{HD} \right]  \Delta s_w^{(4,1)} \nonumber  \\
&+ \frac{2}{s_w^2}\left[ - s_w^2 \Delta M_W^{{\rm eff}(4,1)} + c_w^2 \Delta s_w^{(4,1)}  \right] \Delta M_W^{{\rm eff}(6,0)} \nonumber \\
&   -   \valpha^2 \frac{c_w}{s_w} \frac{1}{\epsilon}\left[ \dot{C}_{HWB}+\frac{c_w}{4s_w}\dot{C}_{HD} \right] \,. 
\label{eq:dV2_veff_NLO}
\end{align}
In the above, we have denoted the tree-level $\alpha$-scheme result as 
\begin{align}
\Delta v_\alpha^{\alpha(6,0)}& =  -2 \frac{c_w}{s_w}
\left[C_{HWB} + \frac{c_w}{4 s_w}C_{HD} \right] \, , 
\end{align}
which leads to the following tree-level results in the  $v_\alpha^{\rm eff}$ scheme:
 \begin{align}
\Delta v_\alpha^{(6,0)} 
 & = -\frac{1}{2}C_{HD} - \frac{1}{c_w s_w}C_{HWB}- \frac{c_{2w}}{c_w^2}\left(g_L^{\ell(6,0)} + \frac{c_{2w}}{2s_w^2}g_R^{\ell(6,0)} \right) \, .
 \end{align}

\begin{table}[t]
	\begin{center}
		\def\arraystretch{1.3}
		\begin{tabular}{|c|c||c|c|}
			\hline $m_t$ & $172.9$~GeV & $v_\mu$                     & 246.2 GeV \\ 
			\hline $M_W$ & $80.38$~GeV & $\valpha$ & 246.5 GeV \\
			\hline $M_Z$ & $91.19$~GeV & $(\sin \theta_{\rm eff}^e)^2$ & 0.23166\\
			\hline  $m_H$     & $125.1$~GeV       & $\alpha(M_Z)$ & 1/128.946 \\
			\hline 
		\end{tabular} 
		\caption{\label{tab:Inputs} Input parameters employed throughout the paper. Note that $v_\alpha$ is a derived parameter. }
	\end{center}
\end{table}

The implementation of the $v_\alpha^{\rm eff}$ scheme also requires to specify the renormalisation scheme for $\alpha$.  One possible 
choice is the \msbar~definition $\overline{\alpha}(\mu) $ in five-flavour QED$\times$QCD, where EW scale contributions are included
through decoupling constants~\cite{Cullen:2019nnr} and perturbative uncertainties can be estimated through scale
variations.  In the current paper we adopt instead the more conventional on-shell definition $\alpha(M_Z)$~\cite{ParticleDataGroup:2022pth}.
It is simple to convert between these two schemes using the perturbative relation 
\begin{align}
\label{eq:Alpha_Convert}
\overline{\alpha}(\mu) &  = \alpha(M_Z)\left[1+ \frac{\alpha(M_Z)}{\pi} \sum_{f\neq t} \frac{N_c^f}{3}Q_f^2\left(\frac{5}{3} +  \ln \frac{\mu^2}{M_Z^2}\right) \right]  \nonumber \\
 & =\alpha(M_Z)\left[1+ \frac{\alpha(M_Z)}{\pi} \left(\frac{100}{27} +  \frac{20}{9}\ln \frac{\mu^2}{M_Z^2}\right) \right]   \, ,
\end{align}
where $Q_f$ is the charge of the fermion and $N_c^f=3$ ($N_c^f =1$)  for quarks (leptons).


\section{Numerical Results}
\label{sec:Numerics}

In this section we present a brief numerical analysis of select electroweak precision observables in the
$v_\sigma^{\rm eff}$ schemes, covering
derived input parameters in section~\ref{sec:GF_MW_Numerics} before turning to heavy EW boson decays in 
section~\ref{sec:W_Z_decay}. We study perturbative convergence and the number of Wilson coefficients associated with 
these schemes, and make qualitative and quantitative comparisons with the widely used $\alpha$, $\alpha_\mu$, and LEP 
schemes.
 
In all cases we use the numerical input parameters given in table~\ref{tab:Inputs}. Furthermore, we approximate the CKM
matrix by the unit matrix. Experimentally, $\sin^2 \theta_{\rm eff}^\ell$ is 
typically averaged over measurements involving electrons and muons. In SMEFT, using an average leads to some difficulties because it would require a combination of first and second-generation Wilson coefficients entering the counterterms, depending on the ratio of electron and muon data entering the combination.  To avoid this issue, we use the most precise available measurement of $\seff$ from the couplings to electrons only, namely the ATLAS measurement with one central and one forward electron~\cite{ATLAS:2018gqq}. 

Results in SMEFT  also depend on the renormalisation scale $\mu$ appearing in the Wilson coefficients $C_i \equiv C_i(\mu)$.  
When variations of this  renormalisation scale are used as a measure of perturbative uncertainties, we use the following 
fixed-order expression for the  RG running,
\begin{align}
\label{eq:C_mu}
C_i(\mu) & = C_i(\mu^{\rm def}) + \ln\left(\frac{\mu}{\mu^{\rm def}}\right)  
 \dot{C}_i(\mu^{\rm def}) \, ,
\end{align}
where $\mu^{\rm def}$ is the default scale for the particular analysis, and the one-loop expressions for $\dot{C}_i$ 
are taken from \cite{Jenkins:2013zja,Jenkins:2013wua,Alonso:2013hga}.
 
\subsection{Derived parameters}
\label{sec:GF_MW_Numerics}
 	
\begin{table}
\centering
\begin{tabular}{ll | c | c | c | c }
 &	& $\frac{M_W^s}{M_W} -1 $ &  $\frac{\alpha^s}{\alpha} - 1$ & $\frac{G_F^s}{G_F} -1 $ & $\frac{\sin^2 \theta_{\text{eff}}^{\ell,s}}{\sin^2 \theta_{\text{eff}}^\ell }   -1 $  \\ \hline
 \multirow{2}{*}{$v_\mu^{\rm eff}$} & LO & $-0.56 \%$  & $0.21 \%$ & -- & --  \\ 
	 & NLO & $0.05 \%$ & $0.23\%$ & -- & -- \\ \hline 
	 \multirow{2}{*}{$v_\alpha^{\rm eff}$} & LO &  $-0.56\%$ &  -- & $-0.21\%$ & --  \\ 
	 & NLO &  $0.04 \%$ & -- & $-0.23\%$   & --  \\ \hline
	 \multirow{2}{*}{$\alpha_\mu$} & LO & --   &  $-2.44 \%$  & -- & $-3.72 \%$   \\ 
	 & NLO & --  & $0.51 \%$  & -- & $0.34 \%$  \\ \hline
	 \multirow{2}{*}{$\alpha$} & LO &  -- & -- &  2.50\%  & $-3.72\%$  \\ 
	 & NLO & --  & -- & $-0.67\%$ & $0.45 \%$   \\ \hline 
	 \multirow{2}{*}{LEP} & LO & $-0.51\%$ & -- & -- & $-0.30 \%$  \\ 
	 & NLO & $0.09 \%$  &-- & --& $-0.32 \%$  \\  \hline \hline	
\end{tabular}
\caption{\label{tab:Derived_SM_rel} SM results for derived parameters in scheme~$s$ relative to the experimental values in table~\ref{tab:Inputs}.}
\end{table}

In each of the five input schemes in table~\ref{tab:schemeDef},  two parameters in the set $\{ M_W, \alpha, \, G_F, \, \seff \}$ are derived parameters which can be calculated in a  SMEFT expansion. For instance, in the $v_\sigma^{\rm eff}$ schemes, the on-shell $W$-boson
mass $M_W$ is given by
\begin{align}
\label{eq:DeltaWdef}
M_W & = M_W^{\rm eff}\left(1+\Delta_W\right) = M_W^{\rm eff}\left(1 +v_\sigma^2 \Delta_W^{(6,0)} + \frac{1}{v_\sigma^2} \Delta_W^{(4,1)} + \Delta_W^{(6,1,\sigma)} \right) \, ,
\end{align}
where $\Delta_W$ is a finite shift. Similarly, the $v_\sigma$ themselves are related according to
\begin{align}
\label{eq:DReff_def}
\frac{1}{v_\mu^2}= \frac{1}{v_\alpha^2}\left[1 +  \Delta r^{\rm eff} \right] 
= \frac{1}{v_\alpha^2}\left[1+ v_\sigma^2 \Delta r^{{\rm eff}(6,0)} + \frac{1}{v_\sigma^2} \Delta r^{{\rm eff}(4,1)} +  \Delta r^{{\rm eff}(6,1)} \right]
\, ,
\end{align}
where in the second equality we made explicit that the expansion coefficients are the same whether $v_\mu$ or $v_\alpha$ is used.  
The form above allows to determine $G_F$ in the $v_\alpha^{\rm eff}$ scheme after setting $\sigma=\alpha$, whereas $\alpha$ in the 
$v_\mu^{\rm eff}$ scheme is easily obtained after setting $\sigma=\mu$.  We derive the SMEFT expansions for $\Delta_W$ and 
$\Delta r^{\rm eff}$, including explicit results in the large-$m_t$ limit, in appendix~\ref{sec:Derived_SMEFT}.  Results for 
$\seff$ in the $\alpha,\alpha_\mu$ and LEP schemes are obtained by evaluating eq.~(\ref{eq:sw_eff_def}), while results for all other 
derived parameters have been given in  \cite{Biekotter:2023xle}.\footnote{We have
converted factors of $\overline{\alpha}(\mu)$ used in that work to the on-shell definition $\alpha(M_Z)$ using eq.~(\ref{eq:Alpha_Convert}).}

The derived parameters are useful for two reasons.  First, from a practical perspective,
they are the simplest examples of EW precision observables and therefore play an important role in global analyses of data.
Second, they are the key ingredients for converting results and understanding differences between
EW input schemes.  For example, if one calculates  a quantity in the $\alpha_\mu$ scheme, one can convert it to the  
$v_\mu^{\rm eff}$ scheme by substituting $M_W$ with eq.~(\ref{eq:DeltaWdef}) with $\sigma=\mu$ and performing a SMEFT expansion. 
In the SM, if the derived value of $M_W$ in the $v_\mu^{\rm eff}$ agrees well with the measured value at a given order, then results for 
other observables in the $\alpha_\mu$ and $v_\mu^{\rm eff}$ scheme will show similar level of numerical agreement. This should 
also be true in SMEFT, but in that case the derived value of $M_W$ depend on the Wilson coefficients, whose values are not precisely known
and are left symbolic.  Differences in observables between schemes show up in non-trivial patterns of Wilson coefficients and the level of 
agreement between schemes is less obvious.  In the remainder of this section we examine derived parameters in the SM across
all schemes, and use the prediction of $M_W$ in the $v_\sigma^{\rm eff}$ schemes to illustrate some of their important  features.

\paragraph{SM.}

In table~\ref{tab:Derived_SM_rel} we show LO and NLO results for derived parameters in the SM, 
where NLO is defined as LO plus the NLO correction.  
In all cases, the NLO and measured values agree to roughly half a percent or better.  In the $v_\sigma^{\rm eff}$ schemes, 
the deviation between the  derived parameters and the experimental values is already below the per-mille level at LO, 
while the $\alpha$ and $\alpha_\mu$ schemes involve percent-level NLO corrections to $\seff$, $\alpha$ or $G_F$.  Such corrections originate mainly from large-$m_t$ corrections  to the counterterm for $s_w$ in those schemes; for instance,  in the $\alpha_\mu$ scheme 
the one-loop SM result is
\begin{align}
\label{eq:SW_universal}
s_{w,0}^2 & = \left( s^{\alpha_\mu}_w\right)^2\left[1 + \left(\frac{c^{\alpha_\mu}_w}{s^{\alpha_\mu}_w}\right)^2 \frac{ \Delta \rho_t^{(4,1)} }{v_\mu^2}  + \dots \right]  \approx
 \left(s_w^{\alpha_\mu}\right)^2\left[1 + 3.3 \% +  \dots \right] \, , 
 \end{align}
where the $\dots$ refers to terms which are subleading in the large-$m_t$ limit, and $c_w^{\alpha_\mu}  = M_W/M_Z$.  
The same result holds in the $\alpha$ scheme  after the replacement $v_\mu \to 2M_W s_w^{\alpha_\mu}/\sqrt{4\pi \alpha}$. 

A  noticeable feature of the $v_\sigma^{\rm eff}$ schemes is that the NLO corrections to $G_F$ or $\alpha$ 
are extremely small.  These corrections are related to $\Delta r^{{\rm eff}(4,1)}$ in eq.~(\ref{eq:DReff_def}), and  
an estimate from the top-loop contribution in eq.~(\ref{eq:RT_IJ}) gives  $-\Delta \rho_t^{(4,1)}/v_\sigma^2\approx -1\%$. 
To understand why this estimate breaks down,  we split the one-loop SM correction into component parts according to 
\begin{align}
\label{eq:GF_SM}
\frac{1}{v_\alpha^2}\Delta r^{{\rm eff}(4,1)} & =\frac{1}{v_\alpha^2} \Delta r_t^{{\rm eff}(4,1)}  +
\frac{1}{v_\alpha^2} \Delta r_{\rm rem}^{{\rm eff}(4,1)} +\frac{\alpha(M_Z)}{\pi}\frac{100}{27}  \nonumber \\
& = \left(-0.9348 + 0.0049 + 0.9143 \right)\% = -0.0156 \% \,,
 \end{align}
where the ordering of the numerical terms on the second line matches those of the analytic expressions above. We note an accidental
cancellation between the large-$m_t$ limit result and that related to the running of $\alpha$ in the on-shell 
scheme\footnote{A similar cancellation occurs in the NLO correction to $\sin^2 \theta^\ell_{\text{eff}}$ in the LEP scheme, 
whose large-$m_t$ correction is obtained 
from eq.~(\ref{eq:SW_universal}) by the replacement $c_w^2/s_w^2\to - c_{2w}/c_w^2$ and is roughly $-1.5\%$ numerically.};  
the latter correction can be  eliminated by converting to the \msbar~definition  as in eq.~(\ref{eq:Alpha_Convert}).
By contrast,  the NLO corrections to $M_W$ in the $v_{\sigma}^{\rm eff}$ scheme 
do not depend on the counterterm for $\alpha$.  The top-loop contribution in eq.~(\ref{eq:MWshift41}) is a good estimate for the NLO correction, as 
seen in the result
\begin{align}
M_{W}^{v_\sigma^{\rm eff}}& =M_W^{\rm eff}\left[1 + \frac{1}{v_\sigma^2}\left( \frac{1}{2}\Delta \rho_t^{(4,1)}+ \Delta_{W,{\rm rem.}}^{(4,1)} \right)  \right] \nonumber \\
& = 79.93 \, {\rm GeV} \, \left[1 + 0.00468  + 0.00137 \right]  = 80.42 \, {\rm GeV} \, , 
\end{align}
where the order of numerical contributions in the second line matches that on the first and we set $v_\sigma=v_\mu$ to obtain 
the numerical value. 
 
\paragraph{SMEFT.}

\begin{figure}
\centering
	\includegraphics[width=0.95\textwidth]{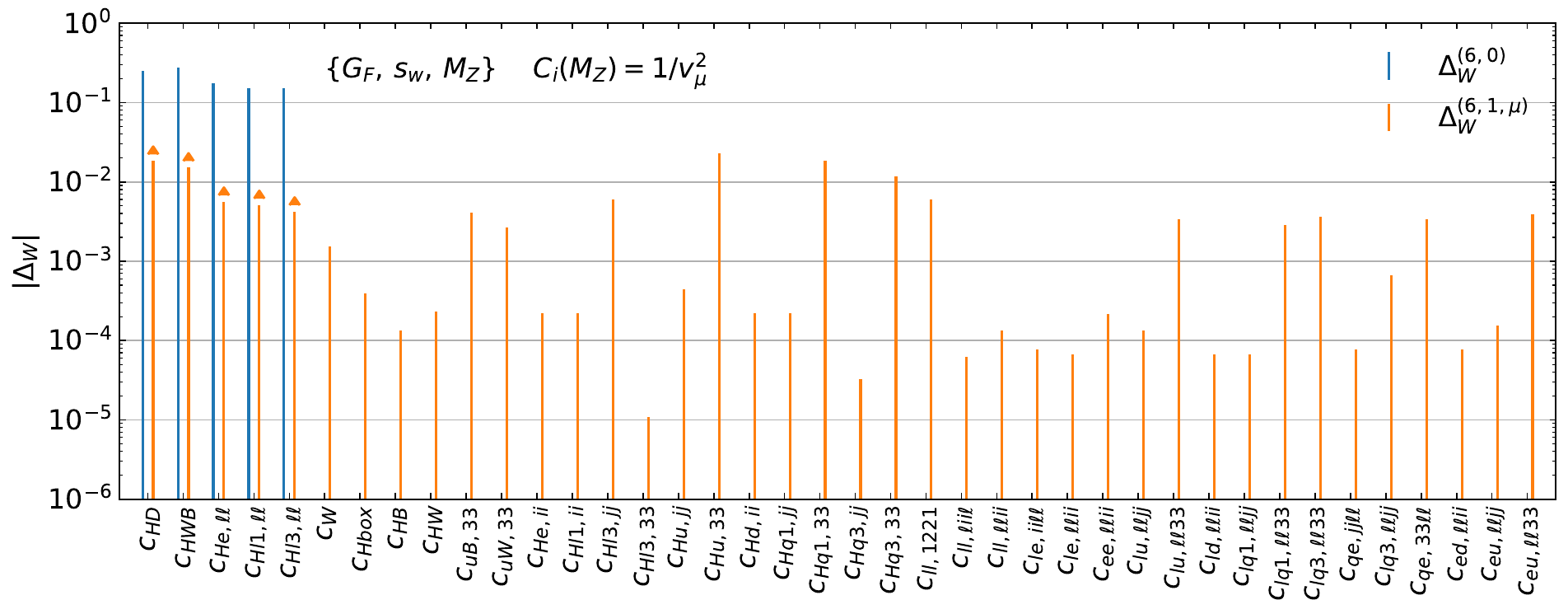}\\
	\includegraphics[width=0.95\textwidth]{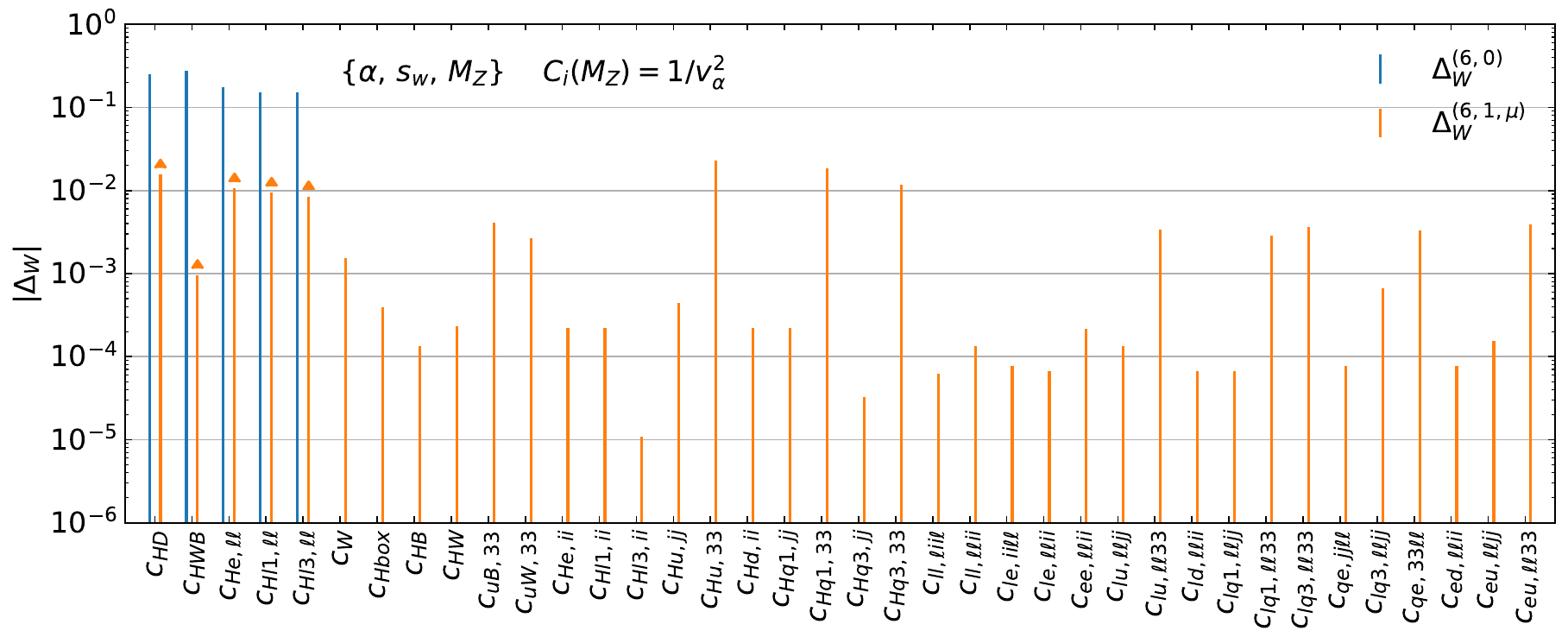}
	\caption{SMEFT corrections to the $W$ boson mass in the $\vmueff$ (top) and $\valphaeff$ (bottom) schemes, with $\Delta s_w$ determined from $Z\to ee$ decay, so that $\ell=1$. The Wilson coefficients are evaluated at $C_i = 1/v_\sigma^2$. The flavour
indices $i$ and $j$ run over values $j \in 1, 2$, and $i \in 1, 2, 3$. }
	\label{fig:DeltaW}
\end{figure}

\renewcommand{\arraystretch}{1.5}
\begin{table}
	\centering
	\begin{tabular}{ll|ccccccc}
		& & $C_{HWB}$ & $C_{HD}$ & $C_{\substack{He\\11}}$ & $C^{(3)}_{\substack{Hl\\11}}$\\ \hline
LO & $v_\sigma^{\rm eff}$ & $-0.275_{-0.009}^{+0.009}$ & $-0.250_{-0.017}^{+0.017}$ & $-0.175_{-0.004}^{+0.004}$ & $-0.151_{-0.003}^{+0.003}$\\
		  \hline

		\multirow{2}{*}{NLO} & $\vmueff$ & $-0.290_{-0.000}^{+0.001}$ & $-0.269_{-0.000}^{+0.003}$ & $-0.180_{-0.000}^{+0.000}$ & $-0.161_{-0.000}^{+0.000}$\\
		 & $\valphaeff$ &  $-0.276_{-0.000}^{+0.000}$ & $-0.266_{-0.000}^{+0.002}$ & $-0.185_{-0.000}^{+0.000}$ & $-0.159_{-0.000}^{+0.000}$\\ \hline
		
		\multirow{2}{*}{$\text{NLO}_t $} & $\vmueff$ & $-0.280_{-0.002}^{+0.003}$ & $-0.261_{-0.003}^{+0.006}$ & $-0.178_{-0.000}^{+0.000}$ & $-0.158_{-0.001}^{+0.001}$ \\
	     & $\valphaeff$ & $-0.272_{-0.002}^{+0.002}$ & $-0.261_{-0.003}^{+0.006}$ & $-0.183_{-0.000}^{+0.000}$ & $-0.158_{-0.001}^{+0.001}$ \\\hline \hline

		& & $C^{(1)}_{\substack{Hl\\11}}$ & $C_{\substack{Hu\\33}}$ & $C^{(1)}_{\substack{Hq\\33}}$& $C^{(3)}_{\substack{Hq\\33}}$ \\ \hline

LO
		 & $v_\sigma^{\rm eff}$ &  $-0.151_{-0.004}^{+0.004}$ &  $0.000_{-0.026}^{+0.026}$ &  $0.000_{-0.026}^{+0.026}$ &   $0.000_{-0.001}^{+0.001}$  \\ 
		 \hline
		
		\multirow{2}{*}{NLO} & $\vmueff$ & $-0.156_{-0.000}^{+0.000}$ & $0.023_{-0.007}^{+0.000}$ & $-0.019_{-0.000}^{+0.006}$ & $0.012_{-0.002}^{+0.000}$\\
		 & $\valphaeff$ & $-0.160_{-0.000}^{+0.000}$  & $0.023_{-0.006}^{+0.000}$ & $-0.019_{-0.000}^{+0.006}$ & $0.012_{-0.002}^{+0.000}$ \\ \hline
		
	\multirow{2}{*}{$\text{NLO}_t $} & $\vmueff$ & $-0.154_{-0.000}^{+0.000}$ & $0.024_{-0.005}^{+0.000}$ & $-0.024_{-0.000}^{+0.005}$& $0.009_{-0.002}^{+0.000}$ \\
		  & $\valphaeff$ & $-0.158_{-0.000}^{+0.000}$  & $0.024_{-0.005}^{+0.000}$ & $-0.024_{-0.000}^{+0.005}$ & $0.009_{-0.002}^{+0.000}$\\\hline \hline
	\end{tabular}
	\caption{\label{tab:DeltaW_vsigma} The numerical prefactors of the Wilson coefficients in the $\vmueff$ and $\valphaeff$ schemes contributing to $\Delta_W$ for the LO, NLO and NLO$_t$ (large-$m_t$ limit) perturbative approximations. The SM tree-level approximation along with $\vmu^2$ has been factored out. The results have been evaluated at $\mu$ = $M_Z$ and varied up and down by a factor of 2 to give the uncertainties. Only Wilson coefficients whose numerical prefactor is greater than 1\% at NLO have been included.}
\end{table}
\renewcommand{\arraystretch}{1}
Results for derived parameters in SMEFT depend on a number of Wilson coefficients and are thus quite lengthy.  For brevity, we 
focus the discussion on $M_W$ in the $v_{\sigma}^{\rm eff}$ schemes, leaving a comparison of observables across
schemes to the heavy-boson decay rates presented in section~\ref{sec:W_Z_decay}.  

We show in figure~\ref{fig:DeltaW}  the LO and NLO SMEFT corrections to $M_W$ in the $v_{\sigma}^{\rm eff}$ schemes.
The numerical contribution from each Wilson coefficient at the scale $\mu=M_Z$ is obtained by making the choice 
$C_i(M_Z) = 1/v_\sigma^2$, and the results are given in units of $M_W^{\rm eff}$; in other words, we are quoting results 
for the expansion coefficients of $\Delta_W$ as defined in eq.~(\ref{eq:DeltaWdef}).  
A remarkable feature is the large number of Wilson coefficients contributing to $M_W$
for arbitrary flavour structure;  the exact number of coefficients at  LO (NLO) is 5 (63) in both schemes. Out of these,  34 Wilson coefficients appearing at NLO correspond to different flavour structures of ten four-fermion operators. 
The number of Wilson coefficients is reduced when we employ additional flavour assumptions. For concreteness, we consider the scenario of minimal flavour violation (MFV) where the top Yukawa~\cite{DAmbrosio:2002vsn} is the only source of the breaking of the $U(3)^5$ symmetry in SMEFT, see appendix~\ref{app:flav_assumptions}.
Under the MFV assumption, 34 Wilson coefficients contribute to $\Delta_W$ at NLO, 16 
of which correspond to different flavour structures of the ten four-fermion operators.   

It is clear from figure~\ref{fig:DeltaW} that many of the NLO SMEFT corrections to $M_W$ are numerically small when 
all Wilson coefficients are set to a common value. In table~\ref{tab:DeltaW_vsigma},  we give numerical results at LO and 
NLO (defined as the sum of LO plus the NLO correction) for those SMEFT operators whose NLO contribution is 
 larger than  $1\%$ for the default choice $C_i(M_Z)=1/v_\sigma^2$. 
All of these coefficients receive NLO corrections from top loops, and to show their significance we give results where
only the large-$m_t$ limit of these corrections is used (NLO$_t$ in the table).  
In each case, we include scale uncertainties obtained by evaluating the prediction for the three scale choices 
$\mu \in [M_Z, \, 2 M_Z , \, M_Z/2]$, using  eq.~(\ref{eq:C_mu}) to express the results in terms of $C_i(M_Z)$. 
In most cases there is a good convergence between LO and NLO when scale uncertainties are included. The
large-$m_t$ limit results are generally an improvement for central values,  but come with small scale uncertainties 
which do not always overlap with the complete NLO result.

\subsection{Heavy boson decays at NLO}
\label{sec:W_Z_decay}

In this section we analyse $W\to \tau\nu$ and $Z\to \tau\tau$ decays, focussing on a comparison between input schemes.
We define SMEFT expansion coefficients for the decay $X\to f_1 f_2$ in input scheme $s$ as 
\begin{align}
\Gamma(X\to f_1 f_2)  & = \Gamma^{s (4,0)}_{Xf_1 f_2} + \Gamma^{s (6,0)}_{Xf_1 f_2}  +  \Gamma^{s (4,1)}_{Xf_1 f_2} + \Gamma^{s(6,1)}_{Xf_1 f_2} \,.
\end{align}
Moreover, we write LO and NLO results as 
\begin{align}
\Gamma^{s}_{Xf_1f_2,\text{LO}} & \equiv \Gamma^{s (4,0)}_{Xf_1 f_2} + \Gamma^{s (6,0)}_{Xf_1 f_2}  \, , \nonumber \\
\Gamma^{s}_{Xf_1f_2,\text{NLO}} & \equiv  \Gamma^{s}_{Xf_1f_2s,\text{LO}} +\Gamma^{s (4,1)}_{Xf_1 f_2} + \Gamma^{s (6,1)}_{Xf_1 f_2}  \,.
\end{align}
The LO results for $s\in\{v_\mu^{\rm eff},v_\alpha^{\rm eff}\}$ are given by  
\begin{align}
 \Gamma^s_{W\tau\nu, \text{LO}} &  = \frac{M_W^{\rm eff} }{12\pi} \left\{   \left(\frac{M_W^{\rm eff}}{v_\sigma}\right)^2 + 
(M_W^{\rm eff})^2 \left[ 2 C_{\substack{Hl \\ 33}}^{(3)}  + 3 \Delta_W^{(6,0)} -\Delta v_\sigma^{(6,0)}   \right]     \right\} \, , \nonumber \\
 \Gamma^s_{Z\tau\tau, \text{LO}} &  = \frac{M_Z}{24 \pi }\bigg\{\frac{M_Z^2}{v_\sigma^2}\left(1- \frac{v_\sigma^2}{2}C_{HD} \right) g^{(4,0)}  
	+M_Z^2\Big[ \left(2\left( g_R^{\ell(6,0)}-g_L^{\ell(6,0)}\right) - \Delta v_\sigma^{(6,0)} \right) g^{(4,0)} \nonumber   \\ 
	&  + 2c_{2w}\left(g_L^{\tau(6,0)}-g_L^{\ell(6,0)}\right) + 4 s_w^2 \left(g_R^{\tau(6,0)}-g_R^{\ell(6,0)}\right) \Big] \bigg\} \, ,
	\label{eq:WZ_decay_LO}
\end{align}
where $\ell\equiv \ell^i$ is the charged lepton species used in the definition of $\Delta s_w$ and 
\begin{align}
\label{eq:gtau40}
g^{(4,0)} &= 1 -4s_w^2 + 8s_w^4  \, .
\end{align}
To derive the result for $W$ decay we have written $W$-mass dependence arising both through two-body phase-space and in the matrix element squared in terms of $M_W^{\rm eff}$.  Note that in $Z$ decay the flavour-independent coupling $G^{(6,0)}$ given in eq.~(\ref{eq:gdef}) has
dropped out of the decay rate due to a cancellation against $\Delta s_w^{(6,0)}$. Further simplifications occur only if $\ell =\tau$ or a flavour
symmetry such as MFV is imposed, in which case the contribution in the final line vanishes.

 \begin{table}
\centering
\begin{tabular}{ll | c | c}
 &	& $\Gamma^s_{W\ell \nu}$/ $\Gamma_{W\ell \nu}^{\text{exp}} -1 $ &  $\Gamma^s_{Z\ell\ell}$/$\Gamma_{Z\ell\ell}^{\text{exp}} -1 $  \\ \hline
 \multirow{2}{*}{$v_\mu^{\rm eff}$} & LO & $-1.30  \%$  &  $-0.70 \%$  \\ 
	 & NLO & $0.16 \%$  & $0.12\%$ \\ \hline 
	 \multirow{2}{*}{$v_\alpha^{\rm eff}$} & LO &   $-1.51  \%$ &   $-0.91  \%$  \\ 
	 & NLO &  $-0.06  \%$ &  $-0.11  \%$ \\ \hline \hline
	 \multirow{2}{*}{$\alpha_\mu$ }& LO & $0.37\%$ & $-0.08  \%$  \\ 
	 & NLO & $0.03 \%$ & $-0.07 \% $\\  \hline	
	\multirow{2}{*}{ $\alpha$} & LO & $2.87\%$  &  $2.41\%$ \\ 	
	& NLO &  $-0.66  \%$  & $-0.74  \%$\\ \hline	
	\multirow{2}{*}{LEP }& LO &  $- 1.17 \%$  &  $-0.66  \%$  \\ 	
	& NLO & $0.31\%$  & $0.16 \%$  \\ \hline \hline	
\end{tabular}
\caption{\label{tab:WZ_decay_SM_rel} Deviations of the SM predictions for $Z\to \ell\ell$ and $W\to \ell\nu$ decay rates in scheme~$s$
from the experimental measurements of $\Gamma_{Z\ell\ell}^\text{exp} = 83.98$~MeV and $\Gamma_{W\ell\nu}^\text{exp} = 226.4$~MeV~\cite{ParticleDataGroup:2022pth}.  
}
\end{table}

	\renewcommand{\arraystretch}{1.5}
\begin{table}
	\centering
	\setlength{\tabcolsep}{2pt}
	\begin{tabular}{ll|ccccc}
		& & $C_{HD}$ & $C_{HWB}$ & $C_{\substack{He\\33}}$ & $C_{\substack{Hu\\33}}$ & $C^{(3)}_{\substack{Hq\\33}}$\\ \hline

		\multirow{2}{*}{$v_\mu^{\rm eff}$} &  LO & $-0.500_{-0.033}^{+0.033}$ & $0.000_{-0.000}^{+0.000}$ & $-1.843_{-0.048}^{+0.048}$ & $0.000_{-0.052}^{+0.052}$ & $0.000_{-0.000}^{+0.000}$\\
		
		&NLO & $-0.527_{-0.000}^{+0.005}$ & $0.004_{-0.000}^{+0.000}$ & $-1.905_{-0.000}^{+0.004}$ & $0.048_{-0.013}^{+0.000}$ & $0.022_{-0.004}^{+0.000}$\\ \hline
		
		\multirow{2}{*}{$v_\alpha^{\rm eff}$} &LO & $0.000_{-0.000}^{+0.000}$ & $2.370_{-0.081}^{+0.081}$ & $-1.843_{-0.050}^{+0.050}$ & $0.000_{-0.003}^{+0.003}$ & $0.000_{-0.005}^{+0.005}$\\
		
		& NLO & $-0.001_{-0.000}^{+0.000}$ & $2.439_{-0.006}^{+0.000}$ & $-1.903_{-0.000}^{+0.004}$ & $0.005_{-0.001}^{+0.000}$ & $0.002_{-0.000}^{+0.000}$\\ \hline \hline
		
		\multirow{2}{*}{$\alpha_\mu$}  &LO & $-0.169_{-0.011}^{+0.011}$ & $0.355_{-0.012}^{+0.012}$ & $-1.764_{-0.046}^{+0.046}$ & $0.000_{-0.018}^{+0.018}$ & $0.000_{-0.001}^{+0.001}$\\
		
		& NLO & $-0.289_{-0.007}^{+0.009}$ & $0.258_{-0.004}^{+0.003}$ & $-1.897_{-0.002}^{+0.006}$ & $0.018_{-0.016}^{+0.011}$ & $0.006_{-0.002}^{+0.000}$\\ \hline 
		
		\multirow{2}{*}{$\alpha$} & LO & $1.573_{-0.108}^{+0.108}$ & $4.088_{-0.143}^{+0.143}$ & $-1.764_{-0.050}^{+0.050}$ & $0.000_{-0.162}^{+0.162}$ & $0.000_{-0.008}^{+0.008}$\\
		
		&NLO & $1.408_{-0.019}^{+0.002}$ & $3.869_{-0.013}^{+0.002}$ & $-1.898_{-0.002}^{+0.006}$ & $-0.142_{-0.000}^{+0.030}$ & $-0.073_{-0.000}^{+0.014}$\\ \hline 
		
		\multirow{2}{*}{LEP}  &  LO & $-0.600_{-0.040}^{+0.040}$ & $-0.474_{-0.016}^{+0.016}$ & $-1.837_{-0.048}^{+0.048}$ & $0.000_{-0.062}^{+0.062}$ & $0.000_{-0.001}^{+0.001}$\\
		
		&NLO & $-0.631_{-0.000}^{+0.005}$ & $-0.475_{-0.000}^{+0.001}$ & $-1.899_{-0.000}^{+0.004}$ & $0.057_{-0.015}^{+0.000}$ & $0.025_{-0.005}^{+0.000}$\\ \hline \hline
	\end{tabular}
	\caption{\label{tab:GammaZ_across_schemes} Selected SMEFT contributions to the $Z\to \tau\tau$ decay rate including scale variation in the five schemes. }
	\end{table}
	\renewcommand{\arraystretch}{1}

 The LO and NLO  results for $W$ and $Z$ decay in the SM are shown in table~\ref{tab:WZ_decay_SM_rel}, 
where we have normalised the results to the experimentally measured values.  
The NLO corrections in the SM bring results from all five schemes into close numerical agreement.   These corrections
are at the $1.5\%$ level for $W$ decay in the $v_\sigma^{\rm eff}$ and LEP schemes, where $M_W$ is not an input and hence factors
of $3\Delta_W^{s(4,1)}/v_\sigma^2$ arise compared to the $\alpha_\mu$ scheme.  
Corrections of around $-3\%$ arise in the $\alpha$ scheme, which are mainly due to the top-loop corrections to $s_w$ 
shown in eq.~(\ref{eq:SW_universal}).  As explained in section~\ref{sec:GF_MW_Numerics}, the close agreement between decay
rates at NLO different schemes is a consequence of that for the derived parameters in table~\ref{tab:Derived_SM_rel}.

The situation in SMEFT is different, because in that case the relations between input parameters in different schemes depend on the
Wilson coefficients and there can be non-trivial interplay with other, process-dependent contributions.  Therefore, in general, the numerical
prefactor multiplying a particular Wilson coefficient can be very different across schemes. 
This point is seen in table~\ref{tab:GammaZ_across_schemes}, where the LO and NLO contributions for an illustrative
sample of  Wilson coefficients are shown for the decay $Z\to \tau\tau$,  using $\ell^i=e$ to determine the $\Delta s_w$ 
counterterm in the $v_\sigma^{\rm eff}$ schemes.  The results include 
perturbative uncertainties obtained by varying the default scale choice $\mu_{\rm def}=M_Z$ up and down by a 
factor of two. We note the following features:
\begin{itemize}
\item The contributions from the coefficients $C_{HD}$ and $C_{HWB}$ have rather different prefactors in each scheme, and convergence
between LO and NLO also differs markedly from case to case -- especially in the $\alpha_\mu$ scheme
the NLO corrections are large and well outside the LO scale uncertainties. 
\item By contrast, at LO the coefficient $C_{\substack{He\\33}}$ appears only in the $Z\to \tau\tau$ matrix element.  The dominant 
NLO corrections arise from SM counterterms on this LO vertex, and tend to push the NLO results in different schemes to similar values.  
The NLO  corrections in the $\alpha$ and $\alpha_\mu$ schemes are outside the LO scale 
uncertainties.
\item The coefficients $C_{\substack{Hu\\33}}$ and  $C^{(3)}_{\substack{Hq\\33}}$ first appear at NLO for fixed $\mu$. The contribution 
of the former is well estimated by LO scale uncertainties through the running of $C_{HD}$ (driven by the top-loop contribution shown
in eq.~(\ref{eq:CGdots})), while that of the latter is unrelated to RG running and requires a genuine NLO calculation. 
\end{itemize}
Regarding the first two points, NLO corrections in the $v_\sigma^{\rm eff}$ schemes tend to be milder than in the $\alpha$ or $\alpha_\mu$ schemes because in the latter case $\Delta s_w$ gets scale-independent corrections of the type shown in eq.~(\ref{eq:SW_universal}). In that 
case including universal corrections from the large-$m_t$ limit using the procedure outlined in  \cite{Biekotter:2023xle} can improve 
convergence between orders.\footnote{A similar procedure could be followed for the $v_\sigma^{\rm eff}$ schemes using eq.~(\ref{eq:RT_IJ}) as a starting point.}  The specific pattern of contributions described above is specific to $Z$ decay, but the important point that 
the size of  SMEFT contributions related to a particular Wilson coefficient is highly scheme specific is general.

\begin{table}[t]
\centering
\begin{tabular}{c c || cc ||  cc ||cc}
&     & \multicolumn{2}{c}{$\Gamma_{W \tau \nu}$} & \multicolumn{2}{c}{$\Gamma_{Z \tau \tau}$ ($\Gamma_{Z ee}$)}  & \multicolumn{2}{c}{Total \# unique WC}  \\ 
                              &     & gen &  MFV  & gen &  MFV  & gen &  MFV  \\ \hline \hline
\multirow{2}{*}{$v_\mu^{\rm eff}$}     & LO  &   8    &         6                
		&  9 (6)   &  4   
		& 10 & 6         \\
                              & NLO &   69    &     34             
                                &    93 (63)    &   33 
                                &    93     &   34        \\ \hline
\multirow{2}{*}{$v_\alpha^{\rm eff}$} & LO  &   6    & 5             &  
		  7  (4)   &     4   &   
		  8 & 5         \\
                              & NLO &    68  &  34    
                               &     92 (63)   &         34   &
                               92 & 34    \\ \hline \hline
\multirow{2}{*}{$\alpha_\mu$}          & LO  &  4      &   1   
               & 8 (7)      &    6 
               & 8       &    6     \\
                              & NLO &   25   &      14     
                               &    67 (64)    &       34   
                                &    67     &       34      \\ \hline
\multirow{2}{*}{$\alpha$}          & LO  & 3      &      3   
           & 5  (5)    &       5  
           & 5      &           5    \\
                              & NLO &   35  &      22    
                               &    63 (63)    &       34  
                               &    63     &       34         \\ \hline
\multirow{2}{*}{LEP}          & LO  &  6      &      4       
         & 8 (7)     &           6   
         & 8      &           6     \\
                              & NLO &   39   &      22  
                                     &    67  (64)   &       34    
                                     &    67    &       34  \\ \hline
\end{tabular}
\caption{Number of Wilson coefficients appearing in heavy boson decay rates under general flavour assumptions and MFV. Note that assuming MFV the number of operators appearing in $\Gamma_{Z\tau\tau}$ and $\Gamma_{Zee}$ is the same and hence only one value is given.  }
\label{tab:Coefficient_Counting}
\end{table}

The discussion so far highlights the non-trivial pattern of perturbative convergence across schemes.  It is also interesting 
to study the number of Wilson coefficients characteristic of each scheme. In table~\ref{tab:Coefficient_Counting}, 
we show the total number of Wilson coefficients contributing to $W$ and $Z$ decay at LO and NLO in the five input schemes for 
different flavour assumptions.  For both decays, significantly more coefficients are involved in the $v_\sigma^{\rm eff}$ schemes than in
the others, when no flavour restrictions are made; this is because the counterterm for $s_w$ is determined from $Z\to ee$ decay amplitudes, 
and involves a number of flavour specific  left and right-handed fermion gauge couplings in addition to four-fermion operators, which
would not otherwise appear in $Z\to \tau\tau$ decay.  Indeed, if we consider the decay $Z\to ee$ instead, the number of Wilson coefficients appearing throughout all five schemes is far more similar.  The same statement applies if MFV is used -- in fact, the number of Wilson coefficients entering the combination of the two decays is the same.  

In a full analysis of electroweak precision observables including gauge-boson decays to quarks and $Z$ decay to neutrinos, the total number of Wilson coefficients appearing is further increased through contributions from four-quark operators.  For MFV the total number of operators appearing grows from 34 in the leptonic $Z$ and $W$ decays considered here to 56 in the full set of electroweak precision observables~\cite{Bellafronte:2023amz}.

\section{Conclusions}
\label{sec:conclusions}

We have implemented to NLO in SMEFT two EW input schemes involving $\seff$ as an input parameter.  These ``$v_\sigma^{\rm eff}$ schemes''
share as common inputs $\seff$ and $M_Z$, but differ through the use of the fine structure constant $\alpha$ ($v_\alpha^{\rm eff}$ scheme) 
or the Fermi constant $G_F$  ($v_\mu^{\rm eff}$ scheme) as the third independent input parameter.  
Details of the renormalisation procedure in these schemes were given in section~\ref{sec:seff}, 
and numerical results for a select set of electroweak precision observables, including comparisons with the other
commonly used EW input schemes listed in table~\ref{tab:schemeDef}, were given in section~\ref{sec:Numerics}.
Analytic results in the $v_\sigma^{\rm eff}$ schemes which form that basis of that numerical analysis 
are given in  electronic form in the arXiv submission of this paper.  

An attractive feature of the $v_\sigma^{\rm eff}$ schemes in SMEFT is that sizeable corrections to the sine of the Weinberg angle 
related to top-quark loops appearing in other schemes are absorbed into the definition of the parameter $\seff$.  
On the other hand, the renormalisation conditions for $\seff$ are implemented at the level of form factors for two-body 
$Z\to \ell \ell$ decay, and are thus subject to a large number of flavour-specific $Z$-fermion SMEFT couplings, including four-fermion
operators.    For instance, the SMEFT expansion for $M_W$ in these schemes receives contributions from five Wilson coefficients at LO, but 63 already at NLO,  and as shown in table~\ref{tab:Coefficient_Counting} even a simple process such as $Z\to \tau\tau$ is subject to roughly 
90 coefficients at NLO.  Flavour assumptions on the SMEFT Wilson coefficients imposed by symmetries such as MFV may therefore
be an essential ingredient to practical implementations of this scheme in global fits to data. 

Regarding such fits, observables in each input scheme are subject to a different pattern of higher-order corrections in both loops 
and operator dimensions. Therefore, performing fits in multiple EW input schemes can provide an important estimate on the 
significance of such missing  corrections, and is a valuable consistency check on the results in any one scheme.  
The results
of this paper provide an important new component for such analyses.

\section*{Acknowledgements}
AB is supported by  the  Cluster  of  Excellence  ``Precision  Physics,  Fundamental
Interactions, and Structure of Matter" (PRISMA$^+$ EXC 2118/1) funded by the German Research Foundation (DFG) within the German Excellence Strategy (Project ID 390831469).  BP would like to thank Marek Sch\"onherr for informative discussions on SM implementations of the input 
schemes studied here.

\appendix  

\section{Expansion coefficients and large-$m_t$ limit of $\Delta_W$ and $\Delta r^{\rm eff}$}
 
\label{sec:Derived_SMEFT}
In this section we derive the  SMEFT expansions of $\Delta_W$ and $\Delta r^{\rm eff}$, which are used 
to predict $M_W$, $G_F$, and $\alpha$ in the  $v_\sigma^{\rm eff}$ schemes. We also give explicit results in the large-$m_t$ limit.

\subsection{SMEFT expansion of $M_W$}
 
 A simple way to derive the quantity $\Delta _W$ defined in eq.~(\ref{eq:DeltaWdef}) 
is to use the bare mass $M_{W,0}$ as an intermediary:
\begin{align}
M_{W,0} = M_W^{\rm eff}\left(1+ \Delta M_W^{\rm eff} \right) = M_W\left(1+\Delta M_W \right) \, .
\end{align}
After expressing the expansion coefficients of the on-shell counterterm $\Delta M_W$  in terms of $M_W^{\rm eff}$ following the notation
of eq.~(\ref{eq:Z61_ren}), one finds
\begin{align}
\label{eq:DW_sigma}
\Delta_W^{(6,0)} & = \Delta M_W^{{\rm eff}(6,0)} \, , \nonumber \\
\Delta_W^{(4,1)} & = \Delta M_W^{{\rm eff}(4,1)} - \Delta M_W^{(4,1)} \, , \nonumber \\ 
\Delta_W^{(6,1,\sigma)} & =  \Delta_W^{(6,1)} -\Delta_W^{(4,1)}\Delta v_\sigma^{(6,0)}+ \Delta_W^{(6,0)}
 \left(\Delta v_\sigma^{(4,1)} - 2 \Delta M_W^{(4,1)}\right) \, , 
\end{align}
with 
\begin{align}
 \Delta_W^{(6,1)}& =  \Delta M_W^{{\rm eff}(6,1)} - \Delta M_W^{(6,1)} +  \Delta_W^{(6,0)}\Delta M_W^{(4,1)} \,.
\end{align}
To derive the large-$m_t$ limit results we first note the SM results
\begin{align}
\label{eq:SM_LMTs}
\Delta s_{w,t}^{(4,1)} &=\Delta e_t^{(4,1)} = 0 \, , \nonumber \\
\Delta v_{\alpha,t}^{(4,1)} & = 2\Delta M_{W,t}^{{\rm eff}(4,1)} =  2\Delta M_{Z,t}^{(4,1)}  \, ,\nonumber \\
\Delta v_{\mu,t}^{(4,1)} & = 2\Delta M_{w,t}^{(4,1)} \, .
\end{align}
One then has for the SM correction to $\Delta_W$ in this limit
\begin{align}
\label{eq:MWshift41}
\Delta_{W,t}^{(4,1)} & = \Delta M_{Z,t}^{(4,1)}  - \Delta M_{W,t}^{(4,1)} = \frac{1}{2}\Delta \rho_t^{(4,1)} \, .  
\end{align}
The result in SMEFT can be written in the form 
\begin{align}
 \Delta_{W,t}^{(6,1,\sigma)} & =  \Delta_{W,t}^{(6,1)} + \Delta_{W,t}^{(4,1)}\left( 2\Delta_W^{(6,0)}  \delta_{\sigma\alpha} - \Delta v_\sigma^{(6,0)}\right) \, ,
\end{align}
where $\delta_{\sigma\alpha}$ is the Kronecker delta.  An explicit calculation shows that 
\begin{align}
\Delta_{W,t}^{(6,1)} & = 
2\Delta_{W,t}^{(4,1)}\left[C^{(3)}_{\substack{Hq\\33}} -\frac{\sqrt{2}M_W^{\rm eff}}{m_t} C_{\substack{uW\\33}} -\frac{1}{2}\Delta_W^{(6,0)}\right]  - v_\sigma^2\Dot{\Delta}_{W,t}^{(6,0)} \ln \frac{\mu}{m_t} \, ,
\end{align}
where the logarithmic dependence is governed by
\begin{align}
 \Dot{\Delta}_{W,t}^{(6,0)} & = -\frac{1}{4} \dot{C}_{HD,t} -\frac{s_w}{2c_w}\dot{C}_{HWB,t} + \frac{s_w^2}{2c_w^2}  \dot{g}_{L,t}^{\ell(6,0)}
+  \frac{c_{2w}}{4c_w^2} \dot{g}_{R,t}^{\ell(6,0)} \, , 
\end{align}
with 
\begin{align}
\label{eq:CGdots}
v_\sigma^2 \dot{C}_{HD,t} & = 8\Delta \rho_t^{(4,1)} \left[C_{HD} + 2 C_{\substack{Hq \\ 33}}^{(1)}  - 2 C_{\substack{Hu \\ 33}}  \right] \, ,\nonumber  \\
v_\sigma^2 \dot{C}_{HWB,t} &=  4\Delta \rho_t^{(4,1)} \left[C_{HWB} -\sqrt{2} \frac{M_Z}{m_t}\left(c_w C_{\substack{uB \\ 33}} +\frac{5}{3} s_w C_{\substack{uW \\ 33}} \right)  \right] \, ,\nonumber \\
v_\sigma^2  \dot{g}_{L,t}^{\ell(6,0)} & = 4\Delta \rho_t^{(4,1)} \left[ g_{L}^{\ell(6,0)} - C^{(1)}_{\substack{lq\\ii33}}+C^{(3)}_{\substack{lq\\ii33}}+
 C_{\substack{lu\\33ii}}\right]\, ,\nonumber \\
v_\sigma^2   \dot{g}_{R,t}^{\ell(6,0)} & = 4\Delta \rho_t^{(4,1)} \left[ g_{R}^{\ell(6,0)} + C_{\substack{eu\\ii33}}-C_{\substack{qe\\33ii}}\right]\, .\
\end{align}

 \subsection{SMEFT expansions of $G_F$ and $\alpha$}
 
The expansion coefficients for $\Delta r^{\rm eff}$ defined in eq.~(\ref{eq:DReff_def}) are calculated similarly to those for $M_W$, except this time using $v_{T,0}$ as an intermediary.   In particular, by equating eq.~(\ref{eq:dV2_mu}) with eq.~(\ref{eq:dV2_alpha})
one finds
 \begin{align}
 \Delta r^{{\rm eff}(6,0)} & = \Delta v_\mu^{(6,0)} - \Delta v_{\alpha}^{(6,0)} \, , \nonumber \\
  \Delta r^{{\rm eff}(4,1)} & = \Delta v_\mu^{(4,1)} - \Delta v_{\alpha}^{(4,1)} \, , \nonumber \\
 \Delta r^{{\rm eff}(6,1)} & = \Delta v_\mu^{(6,1)} - \Delta v_{\alpha}^{(6,1)}  + 2 \Delta v_\mu^{(4,1)}\Delta r^{{\rm eff}(6,0)} \, .
 \end{align}
Following the discussion of universal corrections in \cite{Biekotter:2023xle}, we write the results in the large-$m_t$ limit in the form
\begin{align}
\label{eq:RT_IJ}
\Delta r_t^{{\rm eff}(i,j)}= K_{\alpha}^{(i,j)} - K_{\mu}^{(i,j)}  \,.
\end{align}
Results for the  $K_\mu$ can be read off from  \cite{Biekotter:2023xle} (after adapting to our notation), while the results for
$K_\alpha$ are new.  The one-loop result in the SM is
\begin{align}
K_\sigma^{(4,1)} & =-\Delta \rho_t^{(4,1)} \delta_{\alpha\sigma} \, . 
\end{align}
The SMEFT answer takes the form
\begin{align}
K_\sigma^{(6,0)} & =-\Delta v_\sigma^{(6,0)} \, , \nonumber \\
K_\sigma^{(6,1)} & =  - \frac{1}{2} v_\sigma^2 \dot{K}_{\sigma,t}^{(6,0)}\ln\frac{\mu^2}{m_t^2} + k_\sigma^{(6,1)} \, .
\end{align}
One has, for the non-logarithmic pieces
\begin{align}
k_{\mu}^{(6,1)} & =  \Delta \rho_t^{(4,1)}\sum_{j=1,2} \bigg[C_{\substack{Hl \\ jj}}^{(3)}  - C^{(3)}_{\substack{lq \\ jj33}}\bigg] \, , \nonumber\\
k_{\alpha}^{(6,1)} & = 2 K_{\alpha}^{(4,1)}\left(K_\alpha^{(6,0)} + C_{\substack{Hq \\ 33}}^{(3)}  \right) \, ,
\end{align}
whereas the dependence on the renormalisation scale $\mu$ is governed by
\begin{align}
v_\sigma^2 \dot{K}_{\mu,t}^{(6,0)} &= -4\Delta\rho_t^{(4,1)} \sum_{j=1,2} \bigg[C_{\substack{Hl \\ jj}}^{(3)}  - C^{(3)}_{\substack{lq \\ jj33}}\bigg] \, , \nonumber \\
v_\sigma^2  \dot{K}_{\alpha,t}^{(6,0)}  & = \frac{1}{2} \dot{C}_{HD,t}+\frac{1}{c_w s_w} \dot{C}_{HWB,t} + 
\frac{c_{2w}}{c_w^2}\left(\dot{g}_{L,t}^{\ell(6,0)} + \frac{c_{2w}}{2s_w^2}\dot{g}_{R,t}^{\ell(6,0)} \right) \,.
\end{align}
All components needed to evaluate the latter result were given in eq.~(\ref{eq:CGdots}).

\section{Minimal flavour violation}
\label{app:flav_assumptions}

The calculations in this work have been performed with no assumptions on the flavour structure of the 
SMEFT operators.  To reduce the number of free parameters, we can make the assumption of minimal flavour violation (MFV). 
In this appendix, we give details on this flavour scenario. 

In the SM, the $U(3)^5$ symmetry for the SM fermions
\begin{align}
U(3)^5 \equiv U(3)_q \times U(3)_l \times U(3)_u \times U(3)_d \times U(3)_e \, ,
\end{align}
is broken only by the Yukawa couplings~\cite{Gerard:1982mm,Chivukula:1987py}. 
The MFV scenario extends this requirement to SMEFT~\cite{DAmbrosio:2002vsn}. Since we consider all fermions except the top quark to be massless, we only thus only allow the breaking of the $U(3)^5$ symmetry by the top Yukawa coupling~$Y_t$. 
In the MFV case, we thus distinguish Wilson coefficients involving the top quark from those involving first and second-generation up-type quarks.

We change from the flavour-general scenario to MFV by making a set of replacements on the Wilson coefficients, see e.g.~\cite{ Bellafronte:2023amz}.  For operators with two flavour indices involving leptons or down-type quarks, we can suppress the flavour indices
\begin{align}
C_{\substack{x\\ ii}} \to   C_x , \quad 
\text{for } C_x \in C_{He}, \, C_{Hl}^{(1)}, \, C_{Hl}^{(3)}, \,  C_{Hd}\, . 
\end{align}
For the Wilson coefficients with two flavour indices involving up-type quark fields, we explicitly distinguish top-quark couplings 
\begin{align}
C_{\substack{x\\ jj}} \to   C_x  \text{ for } j \in 1, \, 2 \, ,  \quad 
C_{\substack{Hu\\ 33}} \to   C_{Ht} , \quad
C^{(1)}_{\substack{Hq\\ 33}} \to   C^{(1)}_{HQ} , \quad
C^{(3)}_{\substack{Hq\\ 33}} \to   C^{(3)}_{HQ} \, .
\end{align}
For $C_{uB}$ and $C_{uW}$ only Wilson coefficients with third-family indices contribute in the first place so no replacement is necessary. 

For four-fermion operators with two different fermion bilinears as well as $C_{ee}$, which is simplified by a Fierz identity, there is a single coefficient contributing under the MFV assumption when no up-type quarks are involved
\begin{align}
C_{\substack{x\\ ii j j }} \to   C_x  \quad 
\text{for } C_x \in C_{ee} , \,  C_{le} ,  \, C_{ld} ,  \, C_{ed} \, .
\end{align}
In Wilson coefficients involving up-type quark fields we distinguish the third generation 
\begin{align}
C_{\substack{x\\ ii j j }} \to   C_x  \text{ for } j \in 1, \, 2 , \quad 
C_{\substack{lq\\ ii 3 3 }}^{(1)} &\to   C^{(1)}_{lQ} ,   \quad 
C_{\substack{lq\\ ii 3 3 }}^{(3)} \to   C^{(3)}_{lQ} ,   \quad 
C_{\substack{lu\\ ii 3 3 }} \to   C_{lt} ,   \quad 
\nonumber\\
C_{\substack{qe\\  j j ii  }} \to   C_{qe}  \text{ for } j \in 1, \, 2 , \quad 
C_{\substack{qe\\  3 3 ii }} &\to   C_{Qe}  \quad \, .
\end{align}
For $C_{ll}$, which involves two fermion currents of the same species and chirality, there are two $U(3)^5$ symmetric combinations, which we distinguish with a prime
\begin{align}
C_{\substack{ll\\ ii j j }} \to  C_{ll}  , \quad 
C_{\substack{ll\\ i j j i}} \to  C_{ll}^\prime , \quad 
C_{\substack{ll\\ i i i i}} \to  C_{ll} + C_{ll}^\prime  \, .
\end{align}

\newpage
\begin{table}
\begin{center}
\small
\begin{minipage}[t]{4.4cm}
\renewcommand{\arraystretch}{1.5}
\begin{tabular}[t]{c|c}
\multicolumn{2}{c}{$1:X^3$} \\
\hline
$Q_G$                & $f^{ABC} G_\mu^{A\nu} G_\nu^{B\rho} G_\rho^{C\mu} $ \\
$Q_{\widetilde G}$          & $f^{ABC} \widetilde G_\mu^{A\nu} G_\nu^{B\rho} G_\rho^{C\mu} $ \\
$Q_W$                & $\epsilon^{IJK} W_\mu^{I\nu} W_\nu^{J\rho} W_\rho^{K\mu}$ \\ 
$Q_{\widetilde W}$          & $\epsilon^{IJK} \widetilde W_\mu^{I\nu} W_\nu^{J\rho} W_\rho^{K\mu}$ \\
\end{tabular}
\end{minipage}
%
\begin{minipage}[t]{2.5cm}
\renewcommand{\arraystretch}{1.5}
\begin{tabular}[t]{c|c}
\multicolumn{2}{c}{$2:H^6$} \\
\hline
$Q_H$       & $(H^\dag H)^3$ 
\end{tabular}
\end{minipage}
\begin{minipage}[t]{4.9cm}
\renewcommand{\arraystretch}{1.5}
\begin{tabular}[t]{c|c}
\multicolumn{2}{c}{$3:H^4 D^2$} \\
\hline
$Q_{H\Box}$ & $(H^\dag H)\Box(H^\dag H)$ \\
$Q_{H D}$   & $\ \left(H^\dag D_\mu H\right)^* \left(H^\dag D_\mu H\right)$ 
\end{tabular}
\end{minipage}
%
\begin{minipage}[t]{2.5cm}
\renewcommand{\arraystretch}{1.5}
\begin{tabular}[t]{c|c}
\multicolumn{2}{c}{$5: \psi^2H^3 + \hbox{h.c.}$} \\
\hline
$Q_{eH}$           & $(H^\dag H)(\bar l_p e_r H)$ \\
$Q_{uH}$          & $(H^\dag H)(\bar q_p u_r \widetilde H )$ \\
$Q_{dH}$           & $(H^\dag H)(\bar q_p d_r H)$\\
\end{tabular}
\end{minipage}

\begin{minipage}[t]{4.7cm}
\renewcommand{\arraystretch}{1.5}
\begin{tabular}[t]{c|c}
\multicolumn{2}{c}{$4:X^2H^2$} \\
\hline
$Q_{H G}$     & $H^\dag H\, G^A_{\mu\nu} G^{A\mu\nu}$ \\
$Q_{H\widetilde G}$         & $H^\dag H\, \widetilde G^A_{\mu\nu} G^{A\mu\nu}$ \\
$Q_{H W}$     & $H^\dag H\, W^I_{\mu\nu} W^{I\mu\nu}$ \\
$Q_{H\widetilde W}$         & $H^\dag H\, \widetilde W^I_{\mu\nu} W^{I\mu\nu}$ \\
$Q_{H B}$     & $ H^\dag H\, B_{\mu\nu} B^{\mu\nu}$ \\
$Q_{H\widetilde B}$         & $H^\dag H\, \widetilde B_{\mu\nu} B^{\mu\nu}$ \\
$Q_{H WB}$     & $ H^\dag \sigma^I H\, W^I_{\mu\nu} B^{\mu\nu}$ \\
$Q_{H\widetilde W B}$         & $H^\dag \sigma^I H\, \widetilde W^I_{\mu\nu} B^{\mu\nu}$ 
\end{tabular}
\end{minipage}
%
\begin{minipage}[t]{5.2cm}
\renewcommand{\arraystretch}{1.5}
\begin{tabular}[t]{c|c}
\multicolumn{2}{c}{$6:\psi^2 XH+\hbox{h.c.}$} \\
\hline
$Q_{eW}$      & $(\bar l_p \sigma^{\mu\nu} e_r) \sigma^I H W_{\mu\nu}^I$ \\
$Q_{eB}$        & $(\bar l_p \sigma^{\mu\nu} e_r) H B_{\mu\nu}$ \\
$Q_{uG}$        & $(\bar q_p \sigma^{\mu\nu} T^A u_r) \widetilde H \, G_{\mu\nu}^A$ \\
$Q_{uW}$        & $(\bar q_p \sigma^{\mu\nu} u_r) \sigma^I \widetilde H \, W_{\mu\nu}^I$ \\
$Q_{uB}$        & $(\bar q_p \sigma^{\mu\nu} u_r) \widetilde H \, B_{\mu\nu}$ \\
$Q_{dG}$        & $(\bar q_p \sigma^{\mu\nu} T^A d_r) H\, G_{\mu\nu}^A$ \\
$Q_{dW}$         & $(\bar q_p \sigma^{\mu\nu} d_r) \sigma^I H\, W_{\mu\nu}^I$ \\
$Q_{dB}$        & $(\bar q_p \sigma^{\mu\nu} d_r) H\, B_{\mu\nu}$ 
\end{tabular}
\end{minipage}
%
\begin{minipage}[t]{5cm}
\renewcommand{\arraystretch}{1.5}
\begin{tabular}[t]{c|c}
\multicolumn{2}{c}{$7:\psi^2H^2 D$} \\
\hline
$Q_{H l}^{(1)}$      & $(H^\dag i\overleftrightarrow{D}_\mu H)(\bar l_p \gamma^\mu l_r)$\\
$Q_{H l}^{(3)}$      & $(H^\dag i\overleftrightarrow{D}^I_\mu H)(\bar l_p \sigma^I \gamma^\mu l_r)$\\
$Q_{H e}$            & $(H^\dag i\overleftrightarrow{D}_\mu H)(\bar e_p \gamma^\mu e_r)$\\
$Q_{H q}^{(1)}$      & $(H^\dag i\overleftrightarrow{D}_\mu H)(\bar q_p \gamma^\mu q_r)$\\
$Q_{H q}^{(3)}$      & $(H^\dag i\overleftrightarrow{D}^I_\mu H)(\bar q_p \sigma^I \gamma^\mu q_r)$\\
$Q_{H u}$            & $(H^\dag i\overleftrightarrow{D}_\mu H)(\bar u_p \gamma^\mu u_r)$\\
$Q_{H d}$            & $(H^\dag i\overleftrightarrow{D}_\mu H)(\bar d_p \gamma^\mu d_r)$\\
$Q_{H u d}$ + h.c.   & $i(\widetilde H ^\dag D_\mu H)(\bar u_p \gamma^\mu d_r)$\\
\end{tabular}
\end{minipage}

\vspace{0.25cm}

\begin{minipage}[t]{4.75cm}
\renewcommand{\arraystretch}{1.5}
\begin{tabular}[t]{c|c}
\multicolumn{2}{c}{$8:(\bar LL)(\bar LL)$} \\
\hline
$Q_{ll}$        & $(\bar l_p \gamma_\mu l_r)(\bar l_s \gamma^\mu l_t)$ \\
$Q_{qq}^{(1)}$  & $(\bar q_p \gamma_\mu q_r)(\bar q_s \gamma^\mu q_t)$ \\
$Q_{qq}^{(3)}$  & $(\bar q_p \gamma_\mu \sigma^I q_r)(\bar q_s \gamma^\mu \sigma^I q_t)$ \\
$Q_{lq}^{(1)}$                & $(\bar l_p \gamma_\mu l_r)(\bar q_s \gamma^\mu q_t)$ \\
$Q_{lq}^{(3)}$                & $(\bar l_p \gamma_\mu \sigma^I l_r)(\bar q_s \gamma^\mu \sigma^I q_t)$ 
\end{tabular}
\end{minipage}
\begin{minipage}[t]{5.25cm}
\renewcommand{\arraystretch}{1.5}
\begin{tabular}[t]{c|c}
\multicolumn{2}{c}{$8:(\bar RR)(\bar RR)$} \\
\hline
$Q_{ee}$               & $(\bar e_p \gamma_\mu e_r)(\bar e_s \gamma^\mu e_t)$ \\
$Q_{uu}$        & $(\bar u_p \gamma_\mu u_r)(\bar u_s \gamma^\mu u_t)$ \\
$Q_{dd}$        & $(\bar d_p \gamma_\mu d_r)(\bar d_s \gamma^\mu d_t)$ \\
$Q_{eu}$                      & $(\bar e_p \gamma_\mu e_r)(\bar u_s \gamma^\mu u_t)$ \\
$Q_{ed}$                      & $(\bar e_p \gamma_\mu e_r)(\bar d_s\gamma^\mu d_t)$ \\
$Q_{ud}^{(1)}$                & $(\bar u_p \gamma_\mu u_r)(\bar d_s \gamma^\mu d_t)$ \\
$Q_{ud}^{(8)}$                & $(\bar u_p \gamma_\mu T^A u_r)(\bar d_s \gamma^\mu T^A d_t)$ \\
\end{tabular}
\end{minipage}
\begin{minipage}[t]{4.75cm}
\renewcommand{\arraystretch}{1.5}
\begin{tabular}[t]{c|c}
\multicolumn{2}{c}{$8:(\bar LL)(\bar RR)$} \\
\hline
$Q_{le}$               & $(\bar l_p \gamma_\mu l_r)(\bar e_s \gamma^\mu e_t)$ \\
$Q_{lu}$               & $(\bar l_p \gamma_\mu l_r)(\bar u_s \gamma^\mu u_t)$ \\
$Q_{ld}$               & $(\bar l_p \gamma_\mu l_r)(\bar d_s \gamma^\mu d_t)$ \\
$Q_{qe}$               & $(\bar q_p \gamma_\mu q_r)(\bar e_s \gamma^\mu e_t)$ \\
$Q_{qu}^{(1)}$         & $(\bar q_p \gamma_\mu q_r)(\bar u_s \gamma^\mu u_t)$ \\ 
$Q_{qu}^{(8)}$         & $(\bar q_p \gamma_\mu T^A q_r)(\bar u_s \gamma^\mu T^A u_t)$ \\ 
$Q_{qd}^{(1)}$ & $(\bar q_p \gamma_\mu q_r)(\bar d_s \gamma^\mu d_t)$ \\
$Q_{qd}^{(8)}$ & $(\bar q_p \gamma_\mu T^A q_r)(\bar d_s \gamma^\mu T^A d_t)$\\
\end{tabular}
\end{minipage}

\vspace{0.25cm}

\begin{minipage}[t]{3.75cm}
\renewcommand{\arraystretch}{1.5}
\begin{tabular}[t]{c|c}
\multicolumn{2}{c}{$8:(\bar LR)(\bar RL)+\hbox{h.c.}$} \\
\hline
$Q_{ledq}$ & $(\bar l_p^j e_r)(\bar d_s q_{tj})$ 
\end{tabular}
\end{minipage}
\begin{minipage}[t]{5.5cm}
\renewcommand{\arraystretch}{1.5}
\begin{tabular}[t]{c|c}
\multicolumn{2}{c}{$8:(\bar LR)(\bar L R)+\hbox{h.c.}$} \\
\hline
$Q_{quqd}^{(1)}$ & $(\bar q_p^j u_r) \epsilon_{jk} (\bar q_s^k d_t)$ \\
$Q_{quqd}^{(8)}$ & $(\bar q_p^j T^A u_r) \epsilon_{jk} (\bar q_s^k T^A d_t)$ \\
$Q_{lequ}^{(1)}$ & $(\bar l_p^j e_r) \epsilon_{jk} (\bar q_s^k u_t)$ \\
$Q_{lequ}^{(3)}$ & $(\bar l_p^j \sigma_{\mu\nu} e_r) \epsilon_{jk} (\bar q_s^k \sigma^{\mu\nu} u_t)$
\end{tabular}
\end{minipage}
\end{center}
\caption{\label{op59}
The 59 independent baryon number conserving dimension-six operators built from Standard Model fields, in 
the notation of \cite{Jenkins:2013zja}.  The subscripts $p,r,s,t$ are flavour indices, and $\sigma^I$ are Pauli
matrices.}
\end{table}
\FloatBarrier

\bibliography{literature}
\bibliographystyle{JHEP.bst}

\end{document}